\documentclass{aa}

\usepackage[switch]{lineno}
\usepackage{graphicx}
\usepackage{txfonts}
\usepackage{hyperref}
\hypersetup{
    colorlinks=true,
    citecolor=blue, 
    linkcolor = blue,
    urlcolor = blue
    }
\usepackage[flushleft]{threeparttable}
\usepackage{float} 
\usepackage{subcaption}
\usepackage[export]{adjustbox}
\usepackage{wrapfig}
\usepackage[nopatch]{microtype}
\usepackage{booktabs}
\usepackage{gensymb}

\newcommand{\kms}{km~s$^{-1}$}
\newcommand{\arcs}{$^{\prime\prime}$}
\newcommand{\HI}{H\,{\sc i}}
\newcommand{\arcm}{$^{\prime}$}   
\newcommand{\rhohi}{\ensuremath{\rho_{\mathrm{H\textsc{i}}}}}
                       
\newcommand{\rhohii} {\ensuremath{\rm \rho_{H_2}\ }} 

\def\aap {{A\&A\,}}

\def\apjs {{ApJS\,}}
\def\apjl {{ApJL\,}}

\begin{document}

   \title{The vertical structure of the stellar disk in NGC~551}

   \author{Harshal Raut\inst{1,2} \thanks{harshal1908.hr@gmail.com}, Narendra Nath Patra\inst{2} \thanks{naren@iiti.ac.in}, Prerana Biswas\inst{3}, Nirupam Roy\inst{4}, Veselina Kalinova\inst{1}, Sergio Dzib\inst{1}, Dario Colombo\inst{5,1}, Vicente Villanueva\inst{6} \and Sebastián F. Sánchez\inst{7}}

   \institute{Max-Planck-Institut für Radioastronomie (MPIfR), Auf dem Hügel 69, 53121 Bonn, Germany \and
            Department of Astronomy, Astrophysics and Space Engineering, Indian Institute of Technology Indore, Indore 453552, India \and Indian Institute of Astrophysics, Bangalore 560034, India \and Department of Physics, Indian Institute of Science, Bangalore 560012, India \and Argelander-Institut f\"ur Astronomie, University of Bonn, Auf dem H\"ugel 71, 53121 Bonn, Germany \and Departamento de Astronom\'ia, Universidad de Concepci\'on, Barrio Universitario, Concepci\'on, Chile \and Instituto de Astronomía, Universidad Nacional Autónoma de México, A.P. 70-264, 04510 México, D.F., Mexico}

   \date{}

\abstract
  {}
  {This paper aims to self-consistently determine the 3D density distribution of the stellar disk in NGC~551 and to utilize it to study the observational signatures of two-component stellar disks (thin and thick) in galaxies.}
  {Assuming that the baryonic disks are in hydrostatic equilibrium, we solved the Poisson-Boltzmann equation to estimate the 3D density distribution in the stellar disk of NGC~551. Unlike in previous studies, we used integral-field spectroscopic observations to estimate the stellar velocity dispersion. A 3D dynamical model of the stellar disk was built using these density solutions and the observed rotation curve. Using this model, we generated simulated surface brightness maps and compared them with observations to verify the consistency of our modeling. Furthermore, the dynamical model was inclined to $90^\circ$ to produce an edge-on surface density map of the galaxy. We further investigated this map by fitting different 2D functions and plotting vertical cuts in a logarithmic scale to infer observational signatures of two-component disks in galaxies.}
  {We estimated the vertical stellar velocity dispersion in NGC~551 using an iterative method and obtained results consistent with the formalism employed in the Disk Mass Survey. Through dynamical modeling of the stellar disk in NGC~551, we produced moment maps, which reasonably matched the observations, indicating consistent modeling. We examined the simulated edge-on model by taking vertical cuts and decomposing them into multiple Gaussian components. We find that an artificial double Gaussian component arises due to the line-of-sight integration effect, even for a single-component disk. This indicates that decomposing vertical intensity cuts into multiple Gaussian components is an unreliable method for identifying multicomponent disks. Instead, an up-bending break, visible in the plot of the vertical cuts in the logarithmic scale for a two-component disk, serves as a more reliable indicator, which is absent in the case of a single-component disk. We performed 2D fitting on the edge-on surface density map using the product of a scaled modified Bessel function (for the radial profile) and a $\rm sech^2$ function (for the vertical profile) to estimate the stellar disk’s structural parameters. We find that these traditional methods systematically underestimate the scale length and flattening ratio of the stellar disk. Therefore, we suggest using detailed modeling to accurately deduce the structural parameters of stellar disks in galaxies.}
  {}

    \authorrunning{Raut et al.}
   \maketitle

\section{Introduction}
The structural and chemical properties of the stars in the Galaxy have been extensively studied to understand the nature of the stellar disks. Several galaxy surveys with stellar spectroscopy have measured precise chemical and kinematic properties of a large number of stars in the Galaxy \citep{2015RAA....15.1095L,2017AJ....154...94M,2018MNRAS.478.4513B,2020AJ....160...82S,2021A&A...649A...1G,2022A&A...666A.120G}. These measurements provided unprecedented coverage of the stellar disk, both in terms of its spatial distribution and its chemical history. Furthermore, these surveys help us understand the evolutionary history of the stellar disk in the Galaxy. The extensive datasets of these surveys have also led to the strengthening of the claim of the existence of a two-component stellar disk in the Milky Way around the solar neighborhood. The stellar disk is found to have a "thin" and a "thick" component \citep[see, e.g.,][]{1982PASJ...34..365Y,1983MNRAS.202.1025G}. For example, \cite{1983MNRAS.202.1025G} studied the star counts in the Galaxy and found that the vertical distribution showed an up-bending break and was better fit by two exponential components rather than one, indicating the existence of a two-component disk. 

More recent studies using Gaia astrometric data \citep{2018A&A...616A...1G} and Apache Point Observatory Galactic Evolution Experiment (APOGEE) spectroscopic data \citep{2018ApJS..235...42A} have estimated the age of thousands of stars \citep[see, e.g.,][]{2021A&A...645A..85M} in the disk of the Galaxy. These datasets have been further used to build machine learning models and predict the age of an even larger number of stars \citep{2019MNRAS.489..176M,2021MNRAS.503.2814C}. They find distinct age distributions consistent with two different populations in thin and thick disks.

Several spectroscopic observations in the solar neighborhood also show a bimodal distribution in the $[\alpha/\text{Fe}]$ abundance \citep{2003A&A...410..527B,2005A&A...433..185B,2006MNRAS.367.1329R,2011ApJ...738..187L}. All these studies strongly indicate the existence of a chemically distinct two-component stellar disk in the Milky Way. The thin disk has a relatively small-scale height, residing close to the midplane, and consists primarily of young, metal-rich stars. In contrast, the thick disk has a larger extent in the vertical direction and is dominated by metal-poor, old stars \citep{2003A&A...410..527B,2005A&A...433..185B,2006MNRAS.367.1329R,2011ApJ...738..187L,2011ApJ...735L..46B,2014A&A...562A..71B,2018AJ....156..126J}. However, the general radial extent of thin and thick disks in galaxies remains inconclusive, due to the different ways in which the disks are defined. For example, in \cite{2011ApJ...735L..46B}, \cite{2012ApJ...751..131B}, \cite{2015ApJ...808..132H,2017A&A...608L...1H}, and \cite{2023ApJ...954..124I}, the definitions of thick and thin disks are considered based on mono-age or mono-abundance population, and the scale lengths of the thick disks are found to be smaller than the scale lengths of thin disks. On the other hand, in studies such as \cite{2003AJ....125.1958L,2008ApJ...673..864J}, the thick and thin disks were defined geometrically, and the thick disks were found to extend farther than the thin disks. In addition, a thick disk is characterized by a high vertical velocity dispersion and a slow systemic rotation around the galactic center compared to the thin disk \citep{2003A&A...398..141S,2005A&A...433..185B,2011ApJ...735L..46B,2011ApJ...738..187L,2013A&A...555A..12K,2017A&A...605A...1R,2022A&A...667A..98R}. 

Several studies have also found the existence of two-component disks in external galaxies. For example, \citet{2006AJ....131..226Y} analyzed the vertical light distribution in 34 galaxies, fitting the $R-\text{band}$ surface brightness profiles from the du Pont 2.5m telescope from the Las Campanas Observatory. They found that a two-disk model (thin and thick) was required to explain the observed brightness distribution. \cite{2018A&A...610A...5C} analyzed the surface brightness of 141 edge-on galaxies using data from the \textit{Spitzer} Survey of Stellar Structure in Galaxies, the S$^4$G survey \citep{2010PASP..122.1397S,2015ApJS..219....3M}. Out of their 141 galaxies, only the surface brightness distributions of 17 galaxies were best fit by a single disk component. In contrast, the 124 galaxies required at least two disks to be fit, eight of which required three disks for a proper fit. Many recent studies \citep{2016A&A...593L...6C,2019A&A...625A..95P,2019A&A...623A..19P,2021MNRAS.508.2458M} have used Integral-Field Spectroscopy data from MUSE \citep{2010SPIE.7735E..08B} to investigate the stellar kinematics and the stellar populations to study thick disks in edge-on galaxies. 

These studies strongly indicate the existence of two-component disks in galaxies, identified at least geometrically, chemically, and kinematically. Nevertheless, observing and measuring 3D density distributions would provide direct confirmation. However, projection effects contaminate such measurements due to line-of-sight integration. For example, even for edge-on galaxies, the volume density distribution can only be determined directly from observations if the scale height does not change with radius, which has been shown to not always be accurate. Several authors have shown theoretically \citep{2002A&A...394...89N,2007ApJ...662..335B} and observationally \citep{2014A&A...569A..13R,2021ApJ...912..106Y} that the scale height of galactic disks increases with radius. Also, in the case of a two-component disk, the scale lengths of the thin and the thick disks could be different \citep{2008ApJ...673..864J,2011ApJ...735L..46B,2012ApJ...752...51C,2012ApJ...759...98C,2016ApJ...823...30B,2017MNRAS.471.3057M,2022MNRAS.513.4130L,2022A&A...667A..98R,2023ApJ...954..124I}, further complicating the interpretation of the surface density distributions in edge-on galaxies.

To tackle these observational issues, many previous studies relied upon theoretical modeling of the galactic disks \citep[see, e.g.,][]{2002A&A...394...89N,2007ApJ...662..335B,2019MNRAS.484...81P,2020MNRAS.499.2063P}. Assuming the galactic disks to be in vertical hydrostatic equilibrium, joint Poisson's equations could be set up and solved to obtain the 3D density distribution. \citet{2021MNRAS.501.3527P} used this modeling method to identify a two-component molecular disk in the galaxy NGC~6946. In hydrostatic modeling, the vertical velocity dispersion is a vital parameter. It determines the vertical pressure, which balances gravity. In earlier studies, the velocity dispersion in the stellar disks was calculated theoretically due to a lack of spectroscopic observations \citep{2020MNRAS.499.2063P} \citep[see also][where observational data is used for the Milky Way]{2002A&A...394...89N}. Several simplified assumptions were employed to calculate the stellar velocity dispersion. For example, the stellar disk was assumed to be in vertical hydrostatic equilibrium solely under its gravity (the contribution from gas and dark matter was neglected). The stellar scale height was considered to be constant across all radii, and it was calculated from a general flattening ratio (FR) \citep{2008AJ....136.2782L}. However, this FR can vary from galaxy to galaxy \citep{2014ApJ...787...24B,2014MNRAS.441..869D,2017MNRAS.464...48P}. These assumptions (constancy of stellar scale height, neglecting gas, dark matter mass, etc.) are not true in general, and a direct determination of the stellar velocity dispersion is desirable to model the galactic disks hydrostatically. We note that these assumptions were adopted only to calculate the stellar vertical velocity dispersion. The 3D density distribution in different disk components was calculated by solving the hydrostatic equilibrium equation using these stellar velocity dispersion values.

A direct measurement of the vertical velocity dispersion in galaxies is challenging, requiring spectroscopic observations across the entire stellar disk. This was not possible until recently, with the advent of integral field unit (IFU) spectroscopic observations. Such IFU-based surveys as ATLAS3D \citep{2011MNRAS.413..813C}, the Calar Alto Legacy Integral Field Area (CALIFA) survey \citep{2012A&A...538A...8S,2014A&A...569A...1W}, the Mapping Nearby Galaxies at APO (MaNGA) survey \citep{2015ApJ...798....7B}, and the Sydney-Australian Astronomical Observatory Multi-object Integral Field (SAMI) survey \citep{2021MNRAS.505..991C} have opened up the opportunity to estimate the velocity dispersion in the stellar disk. These surveys provide direct measurements of the stellar velocity dispersion and can be used to estimate the realistic 3D density distribution by solving Poisson's equations.

In this study, we analyzed IFU data for the galaxy NGC~551 from Data Release 3 (DR3) of the CALIFA survey \citep{2016A&A...594A..36S} to estimate the vertical velocity dispersion within its stellar disk. We assumed that the disk is in vertical hydrostatic equilibrium under the gravitational influence of the gas disks and the dark matter halo, and we self-consistently solved the coupled Poisson equations to derive a 3D density distribution. Incorporating this density profile along with a rotation curve obtained from \HI~data, we constructed a dynamical model of the stellar disk. This model was then projected onto the sky plane—adjusted for the observed inclination—and convolved with the telescope’s point spread function (PSF) to generate a model surface density map that can be directly compared with observations. Furthermore, by reorienting the model to an edge-on view ($90^\circ$), we obtained an edge-on surface density map of the galaxy, which we analyzed by fitting various 2D functions and examining logarithmic vertical profiles to investigate the signatures of two-component stellar disks.

\section{NGC~551}
\label{sec:Sample}

We performed hydrostatic modeling (as described in the next section) to investigate the 3D structure of the stellar disk in NGC~551. For this galaxy, all the required data are available publicly. Moreover, NGC~551 shows a remarkable similarity with the Milky Way in several aspects, including the morphology of the stellar disk. NGC~551 is classified as an SBbc \citep{1991rc3..book.....D} galaxy, whereas the Milky Way is classified as an SBc \citep[see, e.g., ][]{1983PASP...95..721H} galaxy. The optical diameter of the Milky Way in $B\ \text{band}$ is $\sim 27$ kpc \citep{1998Obs...118..201G}, whereas that of NGC~551 is $\sim 31$ kpc \citep{2017ApJ...846..159B} (the assumed angular diameter distance is 72 Mpc; \citealt{2017ApJ...846..159B}). The SDSS $r-\text{band}$ luminosity of the Galaxy is measured to be $\rm 3.8^{+1.54}_{-1.12} \times 10^{10} \: L_\odot$ \citep{2015ApJ...809...96L}, taking the dimensionless Hubble parameter $h = 0.69$ (where $h = \frac{H_0}{100\rm\:km\: s^{-1}\:Mpc^{-1}}$ with $H_0$ in $\rm km\: s^{-1}\:Mpc^{-1}$) \citep{2021ApJ...919...16F} and $M_{r,\odot}$ = 4.65 \citep{2018ApJS..236...47W}, whereas that of NGC~551 is estimated to be $ \rm 2.0 \times 10^{10} \: L_\odot$ (obtained from the SDSS DR14 survey done by \citealt{2018ApJS..235...42A}). Furthermore, the star formation rate for our Galaxy is $ \rm 2.0 \pm 0.7 \  M_\odot \thinspace yr^{-1}$ \citep{2022ApJ...941..162E}, and that of NGC~551 is $\rm 2.04 \pm 0.33 \ M_\odot \thinspace yr^{-1}$ \citep{2017ApJ...846..159B}. The stellar disk mass of the Galaxy is measured to be $\rm 5.17^{+1.11}_{-1.11} \times 10^{10} \: M_\odot$ (without bulge; \citealt{2015ApJ...806...96L}), whereas that of NGC~551 is measured to be $\rm 5.75 \times 10^{10} \: M_\odot$, which we obtained from integrating the stellar surface density ($\Sigma_{\rm \star}$) profile (seen in Fig.~\ref{fig:SD}). Given these facts, the optical disk of NGC~551 was expected to be very similar to that of the Milky Way, and hence investigating the structure of its stellar disk would provide a representative picture of the multicomponent stellar disks in massive galaxies.

\begin{table}
\begin{threeparttable}
        \centering
        \caption{Basic properties of NGC~551.}
        \begin{tabular}{ lccr } 
 \hline
 Parameters & Values \\
 \hline
$\text{RA}^a$ & $ \rm 01\: 27\: 40.66\:(h\:m\:s) $ \\
$\text{Dec}^a$ & $\rm  +\:37\:10\:58.5\:( $\degree\:\arcm\:\arcs $ ) $  \\
$\text{Inc}^b$ & $63 \degree$ \\ 
$\text{PA}^b$ & $135 $\degree \\
$\text{Dist}^c$ & 74.5 Mpc \\
$\text{SFR}^c$ & $\rm 2.04 \pm 0.33\: M_\odot\:yr^{-1}$ \\ 
$\text{Optical Diameter}^c$ ($\rm D_{25}$) & $89 $\arcs \\
$\text{Stellar Disk Mass without bulge}^d$ ($\rm M_\star$) &                       $\rm 5.75 \times\: 10^{10}\: \rm M_\odot$\\
$\text{Classification}^e$ & SBbc \\
 \hline 
\end{tabular}
\begin{tablenotes}
        \item[a]{\cite{2006AJ....131.1163S}}
        \item[b]{Obtained by performing MGEfit (see Sect.~\ref{sec:surf}) of the galaxy.}
        \item[c]{The luminosity distance obtained from \cite{2017ApJ...846..159B}}.
        \item[d]{Obtained from integrating the stellar surface density ($\Sigma_{\rm \star}$) profile from 2 kpc outward, excluding the inner region.}
        \item[e]{\cite{1991rc3..book.....D}}
    \end{tablenotes}
 \label{tab:1}
\end{threeparttable}
\end{table}

\section{Modeling the galactic disk}

We adopted a similar procedure to model the galactic disk as used by several previous studies \citep{2018MNRAS.478.4931P,2020MNRAS.499.2063P}. We modeled the galaxy disk as a multicomponent system consisting of stellar, molecular, and atomic disks under the external force field of the dark matter halo. We assumed that all these disks are in vertical hydrostatic equilibrium under the combined gravity of all the components, including dark matter. The vertical velocity dispersion of a disk component solely determines the pressure in the vertical direction. This pressure balances the combined gravitational force due to all disk components. For a simplified mathematical description, we further assumed the galactic disks are concentric, coplanar, and axisymmetric. The center of the disks coincides with the center of the dark matter halo. Under these assumptions, Poisson's equation of hydrostatic equilibrium can be expressed in cylindrical coordinates as

\begin{equation}
    \frac{1}{R}\frac{\partial}{\partial R}\biggl(R\frac{\partial\Phi_{\rm tot}}{\partial R}\biggr) + \frac{\partial^2\Phi_{\rm tot}}{\partial z^2} = 4 \pi G \biggl(\sum_{i=1}^{n=3} \rho_i + \rho_{\rm dm}\biggr),
    \label{eq:equation 1}
\end{equation}
\noindent where $\Phi_{\rm tot}$ denotes the overall potential resulting from the gravitational force of the dark matter halo and the baryonic disks. The variable $\rho_i$ denotes the volume density of different disks, with $i=1,2,3$ representing the stellar, the atomic, and the molecular disks, respectively.  $\rho_{\rm dm}$ is the density of the dark matter halo, which is fixed and determined by mass modeling.

In vertical hydrostatic equilibrium, the vertical gradient of the gravitational potential is balanced by the pressure gradient. This fundamental principle can be expressed through the Boltzmann equation. For an elemental volume, it can be written as
\begin{equation}
    \frac{\partial}{\partial z} \Bigl(\rho_i\bigl<\sigma^2_z\bigr>_i\Bigr) + \rho_i \frac{\partial\Phi_{\rm tot}}{\partial z} = 0. 
    \label{eq:equation 2}
\end{equation}

\noindent The symbol $\sigma_z$ denotes the vertical velocity dispersion and was assumed to be independent of $z$ and only a function of radius. By utilizing Eq.~\ref{eq:equation 2}, we rewrote Eq.~\ref{eq:equation 1} as

\begin{equation}  
\begin{split}
    \bigl<\sigma^2_z\bigr>_i \frac{\partial}{\partial z} \biggl(\frac{1}{\rho_i}\frac{\partial\rho_i}{\partial z} \biggr) = & -4 \pi G \Bigl(\rho_\star + \rhohi + \rhohii + \rho_{\rm dm} \Bigr)   \\
    & + \frac{1}{R}\frac{\partial}{\partial R}\biggl(R\frac{\partial\Phi_{\rm tot}}{\partial R}\biggr),
   \end{split}
   \label{eq:equation 3}
\end{equation}

\noindent where $\rho_\star$, $\rhohi$, and $\rhohii$ denote the volume densities of stars, atomic gas, and molecular gas, respectively. The last term on the right-hand side can be estimated using the rotation curve.

\begin{equation}
    \biggl(R\frac{\partial\Phi_{\rm tot}}{\partial R}\biggr)_{R,z} = (V^2_{\rm rot})_{R,z}.
    \label{eq:equation 4}
\end{equation}

\noindent Here, $V_{\rm rot}$ represents the galaxy's rotation velocity. As the rotation curve of the inner region of the galaxy is not flat, we cannot neglect this term, unlike in some previous studies \citep{2002A&A...394...89N,2008ApJ...685..254B,2018A&A...610A...5C}. We assumed that this rotation velocity does not change as a function of height. Consequently, we kept the last term in Eq.~\ref{eq:equation 3} constant at a particular radius while solving the following hydrostatic equilibrium:

\begin{equation}
\begin{split}
    \bigl<\sigma^2_z\bigr>_i \frac{\partial}{\partial z} \biggl(\frac{1}{\rho_i}\frac{\partial\rho_i}{\partial z} \biggr) = & -4 \pi G \Bigl(\rho_\star + \rhohi+ \rhohii + \rho_{\rm dm} \Bigr) \\
    & +  \frac{1}{R}\frac{\partial}{\partial R} \Bigl(V^2_{\rm rot}\Bigr).
\end{split}
     \label{eq:equation 5}
\end{equation}

\noindent Equation ~\ref{eq:equation 5} represents a system of three coupled second-order partial differential equations in $\rho_\star(z)$, $\rhohi(z)$, and $\rhohii(z)$. Solving these equations at different radii would provide the density of different disk components (stars, atomic gas, and molecular gas) as a function of radius and height. This would provide a complete 3D density distribution of a disk.

To solve Eq. \ref{eq:equation 5}, we require several input parameters. For instance, at any given radius, the velocity dispersion values for the individual disk components must be known (first term of the left-hand side of Eq.~\ref{eq:equation 5}). Additionally, we need dark matter halo parameters to estimate the dark matter density distribution ($\rm \rho_{dm}$; first term on the right-hand side). Furthermore, the last term on the right-hand side is computed using the rotation curve. While the initial densities of the disk components used in solving Eq.~\ref{eq:equation 5} can be arbitrary, the solution, i.e., $\rho_i (z)$, must yield the observed surface density when integrated over $z$. Thus, surface densities serve as another key input parameter. In the following subsection, we describe how these input parameters were estimated for NGC~551.

\subsection{Surface densities}
\label{sec:surf}

\begin{figure}
        \includegraphics[width=\columnwidth]{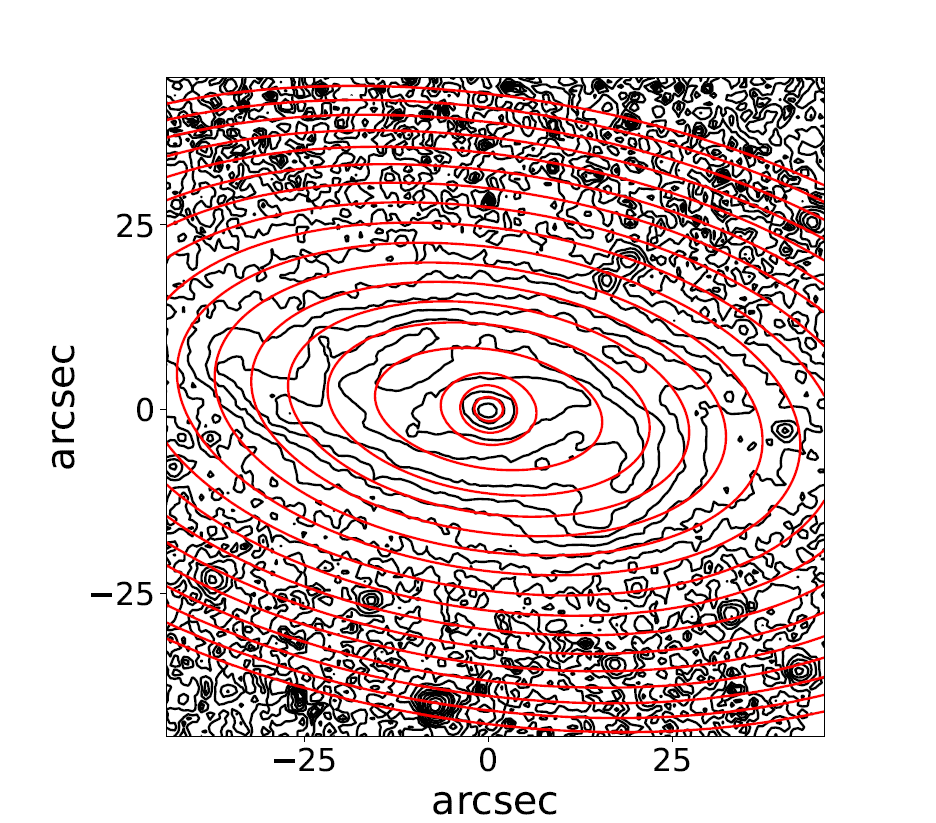}
    \caption{MGE fit of the \textit{Spitzer} 3.6$\,\mu{\rm m}$ image of NGC~551. The black contours represent the observed stellar surface brightness distribution, whereas the red contours indicate the MGE-fitted isophotes. See the text for more details.}
    \label{fig:MgeFit}
\end{figure}

We used the 3.6\,$\mu{\rm m}$ infrared (IR) data (with 680 pc angular resolution at a distance of 72 Mpc) from the \textit{Spitzer} observatory to estimate the surface density profile of the stellar disk. The IR traces the old stars more accurately than optical bands, such as the SDSS $i\ \text{band}$ or $r\ \text{band}$. As a significant amount of mass is locked in old stars in galaxies \citep{2010PASP..122.1397S}, 3.6\,$\mu{\rm m}$ data are suitable for tracing the population of stars that contribute most to the gravity of the disk. We converted the 3.6\,$\mu{\rm m}$ flux into stellar masses using the calibration relation given as $ M_\star = 10^{5.97} \times F_{3.6} \times \bigl( \frac{D_{\rm L}}{0.05} \bigr)^2 $ \citep{2012AJ....143..139E}, where $ F_{3.6}$ is the flux in units of $\rm Jy$, $D_{\rm L}$ is the luminosity distance in units of Mpc, and $ M_\star$ is the stellar mass in units of solar mass ($\rm M_\odot)$.

Estimating the surface density profiles often involves fitting isophotes to the starlight distribution. This fitting technique usually produces good results when the brightness distribution does not deviate from an elliptical shape. However, these deviations are common when a disk consists of multiple components, such as a nucleus, bulge, ring, or spiral arms, or has asymmetries. In these cases, it is better to model the surface brightness as the sum of multiple 2D Gaussians. We used the multi-Gaussian expansion (MGE) technique \citep{1992A&A...253..366M,1994A&A...285..723E,2002MNRAS.333..400C} to estimate the surface density profile of the stellar disk in NGC~551. 

\begin{figure}
        \includegraphics[width=\columnwidth]{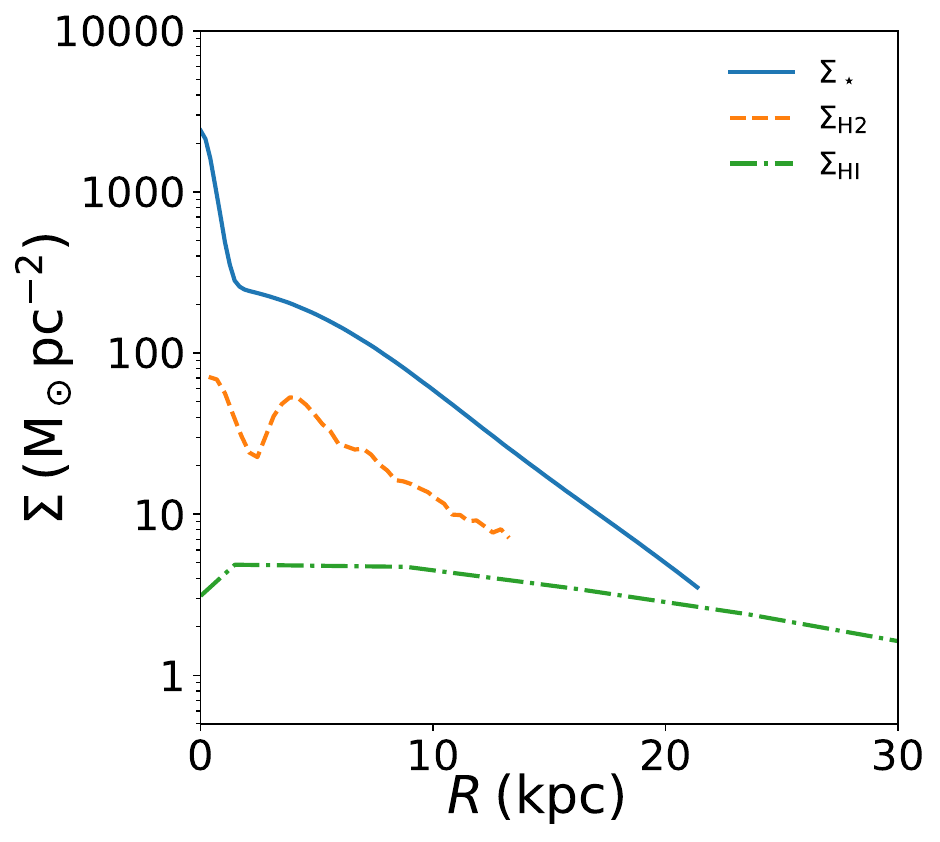}
    \caption{Deprojected surface density profiles of different disks in NGC~551. The solid blue line represents the stellar surface density, whereas the dashed orange and dashed-dotted green lines represent the molecular and atomic gas surface density profiles, respectively. The stellar surface density profile was obtained using the \textit{Spitzer} 3.6\,$\mu{\rm m}$ map and MGE fitting. The $\rm \Sigma_{H2}$ is measured from maps from \citet{2017ApJ...846..159B}. The $\rm \Sigma_{HI}$ profile is obtained through a 3D tilted ring model fitting of the \HI~spectral cube obtained through the uGMRT \citep{1991CSci...60...95S,2017CSci..113..707G} observation. See the text for more details.}
    \label{fig:SD}
\end{figure}

In the MGE method, the galaxy's light distribution is fit with multiple 2D Gaussian components iteratively until the residuals are consistent with background noise. The method accounts for the PSF during the fitting process, ensuring accurate characterization of the galaxy's intrinsic light distribution. The fitted Gaussian components are appropriately deprojected, incorporating the fitted intrinsic axial ratio of the disk to construct the deprojected surface density profile. In Fig.~\ref{fig:MgeFit}, we show the results of the MGE fit. As shown, the fitted model compares well with the isophotes. The corresponding deprojected surface density profile is shown in Fig.~\ref{fig:SD} (blue line).

\begin{figure}
        \includegraphics[width=\columnwidth]{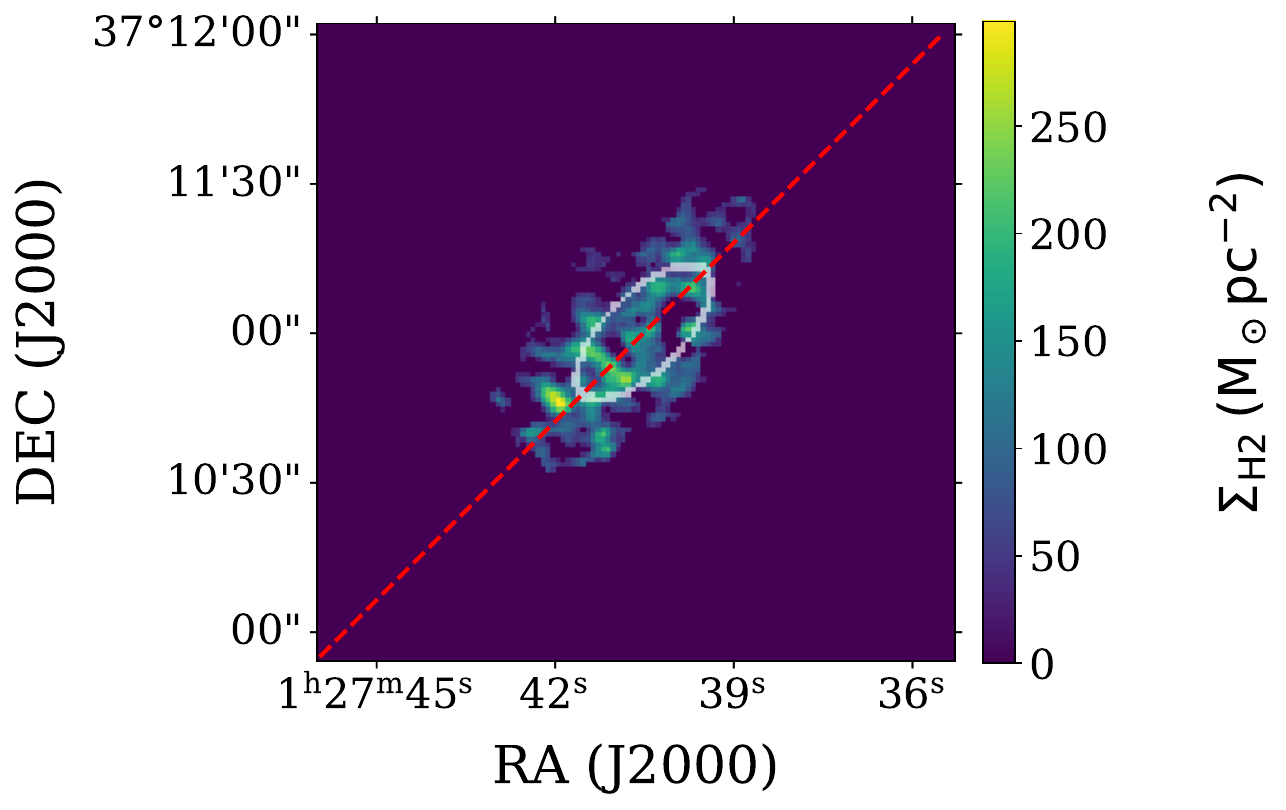}
    \caption{$\rm \Sigma_{H2}$ distribution in NGC~551. The white ellipse shows a representative annular region (a radial bin) at a radius of 6 kpc over which $\rm \Sigma_{H2}$ was averaged to obtain a data point in the $\rm \Sigma_{H2}$ profile. The $\rm \Sigma_{H2}$ map was taken from \citet{2017ApJ...846..159B}. The annular ellipses were demarcated, assuming an inclination of $63\degree$ and a position angle of $135\degree$. The dashed red line designates the major axis of the galaxy.}
    \label{fig:annuli}
\end{figure}

To obtain the surface density profile of the molecular disk in NGC~551, we used molecular gas maps from \citet{2017ApJ...846..159B}. They used the Combined Array for Millimeter-wave Astronomy (CARMA) \citep{2006SPIE.6267E..13B} interferometer to map molecular gas ($\text{CO} \ (J = 1 - 0)$) in 126 nearby galaxies as part of the EDGE-CALIFA survey. We used the total intensity map (the moment zero map with smooth masking) of the molecular gas to estimate the surface density profile. To convert the CO flux into molecular mass (in solar masses), we used the conversion given by \citet{2017ApJ...846..159B},
\begin{equation}
    M_{\text{mol}} = 1.05 \times 10^4 \left(\frac{S_{\text{CO}}\Delta v\: D_{\rm L}^2}{1+z}\right),
\end{equation}
where $S_{\text{CO}}\Delta v$ is the integrated $\ \text{CO} \ (J = 1 - 0)$ line flux (in Jy km s$^{-1}$), $z$ is the redshift. Adopting the inclination and the position angle of the molecular disk from Table \ref{tab:1}, we constructed elliptical annular regions of size $2$\arcs$\:$ at different radii and estimated the surface density profile. We show one such region in Fig.~\ref{fig:annuli}. We note that the gaseous disk in NGC~551 does not show complicated multicomponent structures (bright nucleus, bulge, etc.). Hence, a simple estimation of surface density within an annular region should suffice. The dashed orange line in Fig.~\ref{fig:SD} shows the surface density profile of the molecular gas in NGC~551. 

Lastly, we extracted the \HI~surface density profile for NGC~551 using the \HI~map obtained through the Upgraded Giant Metrewave Radio Telescope (uGMRT) \citep{1991CSci...60...95S,2017CSci..113..707G} observation (proposal code $39\_037$) as part of the Mass modeling and Star Formation Quenching of Nearby Galaxies (MasQue; PI: V. Kalinova; Kalinova et al., in preparation) survey—an interferometric \HI~follow-up of CALIFA galaxies within the EDGE-CALIFA collaboration. The uGMRT data were analyzed using the Astronomical Image Processing Software (AIPS) \citep{2003ASSL..285..109G}, adopting the same approach as described in \citet{2022MNRAS.513..168B}. Consequently, an \HI~spectral cube was produced. A 3D tilted ring model was fit to the spectral cube using Fully Automated TIRIFIC (FAT) \citep{2007A&A...468..731J,2015MNRAS.452.3139K} to obtain the rotation curve (see Appendix B for more details). In addition, we obtained the \HI~surface density profile of the atomic disk, which was multiplied by a factor of 1.33 to account for the helium contribution. The observed \HI~flux was converted to \HI~mass using the formalism given in \citet{2022MNRAS.513..168B}, 
\begin{equation}
    \left(\frac{M_{\text{H\,I}}}{h_{\text{C}}^{-2}\text{M}_{\odot}}\right) \simeq \frac{2.35 \times 10^5}{1+z}\left(\frac{D_{\rm L}}{h_{\text{C}}^{-1}\:\text{Mpc}}\right)^2\left(\frac{S^{V_{\text{rest}}}}{\text{Jy km s}^{-1}}\right),
\end{equation}
where $h_c = \frac{H_0}{100\rm\:km\: s^{-1}\:Mpc^{-1}}$ and $S^{V_{\text{rest}}}$ is the integrated line flux calculated in the rest frame of the galaxy. In Fig.~\ref{fig:SD}, the dashed-dotted green line represents the surface density profile of the atomic disk. As shown in the figure, the stellar surface density profile dominates both the gas disks at all radii. Thus, the stellar disk is expected to primarily determine the vertical density of the stars.

\subsection{Velocity dispersion}
\label{sec:VD}

The velocity dispersion, which directly governs the vertical pressure, is a fundamental parameter in solving the hydrostatic equilibrium equation. However, accurately measuring the velocity dispersion in stellar disks has been challenging due to the limited spatial coverage of long-slit spectroscopy. The advent of IFU observations has particularly advanced this field by enabling measurements of stellar spectra across the entire disk.

In most previous studies on hydrostatic modeling \citep{2019MNRAS.484...81P,2020MNRAS.499.2063P}, the stellar vertical velocity dispersion was calculated analytically while solving the hydrostatic equilibrium equation. These calculations often generalize several properties of the stellar disk. For example, they assume the stellar disk is isothermal and in hydrostatic equilibrium solely under its own gravity \citep{2021MNRAS.501.3527P} \citep[see also][where observational data is used for a non-isothermal stellar disk in the Milky Way]{2020MNRAS.499.2523S}. Furthermore, they adopt a constant stellar scale height \citep{2008AJ....136.2782L} calculated using the observed FR in edge-on galaxies, i.e., $\frac{l_{\rm \star}}{h_{\rm \star}} = 7.3 \pm 2.2$ \citep{2002MNRAS.334..646K}. Here, $l_{\rm \star}$ and $h_{\rm \star}$ represent the scale length and scale height of the stellar disk, respectively. Under these assumptions, the vertical velocity dispersion in a stellar disk can be given as 
\begin{equation}
    \sigma_{\rm \star} = \sqrt{\frac{2 \pi G l_{\rm \star}}{7.3} \Sigma_{\rm \star}}\:\rm km s^{-1},
\end{equation}
where $l_\star$ and $\Sigma_\star$ are in units of kpc, and $\rm M_\odot pc^{-2}$, respectively \citep{1988A&A...192..117V,2008AJ....136.2782L}. The velocity dispersion $\sigma_{\rm \star}$ can be rewritten as 
\begin{equation}
    \sigma_{\rm \star}=1.924 \sqrt{l_{\rm \star} \Sigma_{\rm \star}}\:\rm km s^{-1}.
    \label{eq:analytical_expressiom}
\end{equation} 

However, the assumptions adopted to obtain the above formula are often not satisfied in galaxies and, thus, can lead to significant errors in the estimated vertical velocity dispersion. For example, the scale height in most galaxies increases with radius \citep{2002A&A...394..883L,2002A&A...394...89N,2007ApJ...662..335B,2012EPJWC..1904006M,2014A&A...569A..13R,2014A&A...567A.106L,2016MNRAS.460L..89K,2023MNRAS.523.3915S}. Furthermore, the FR is found to correlate with central surface brightness \citep{2002A&A...389..795B} and morphological type \citep{2014ApJ...787...24B}, both of which are probably correlated with the galaxy's overall luminosity. While deriving the analytical expression, the hydrostatic equilibrium of the disk is considered, but only the gravity of the stellar disk is taken into account (through $\Sigma_\star$). Nevertheless, the dark matter halo and the gas disks can contribute significantly to the system's gravity. This would indeed alter the estimation of the $\sigma_\star$ considerably. Therefore, a direct determination of the $\sigma_\star$ through observation is crucial. For NGC~551, we used the IFU spectroscopic observations from the CALIFA survey to estimate the velocity dispersion in the stellar disk. We used the pyPipe3D pipeline \citep{2022NewA...9701895L}, an enhanced version of the earlier Pipe3D pipeline \citep{2016RMxAA..52..171S}, to process the IFU data. We used the V500 observation setup spanning the wavelength range from 3745 \AA $\:$ to 7300 \AA, with a spectral resolution (R) of $\sim$ 850 $\:$ at 5000 \AA. This setup ensures that the analysis includes stellar absorption lines in the optical band.

The pyPipe3D pipeline takes the IFU data and spatially bins it using the continuum segmentation binning (CS-binning) algorithm \citep{2016RMxAA..52..171S}. The spectra corresponding to the spaxels within each spatial bin were averaged and stored as a single spectrum, together with the average spatial coordinates. Thus, for each bin, we obtained a spectrum that corresponds to the mean of the individual spectra of all the spaxels within that spatial bin, masking spectral pixels with bad values. At the end of this process, the row-stacked spectra (RSS) file was created.
Each spectrum within the RSS file was analyzed using the pyFIT3D spectral fitting code, which separates the stellar continuum and the ionized emission spectrum from the observed spectral energy density for each spaxel. First, the stellar kinematics and dust attenuation were estimated by generating model spectra using a simple set of single stellar populations (SSPs) \citep{2013A&A...557A..86C} convolved with the instrumental dispersion to match the spectral resolution of the data. Next, the ionized gas emission line spectrum was obtained by subtracting the model spectra from the original data. This spectrum was then fit with a few strong, well-known emission lines to obtain the emission line parameters. Next, by subtracting the emission line spectrum from the observed spectrum, we obtained a gas-free spectrum on which stellar population analysis is performed and where the spectrum is decomposed into a larger set of SSPs to obtain different properties such as stellar age and metallicity. In the end, we obtained the final emission-line free cube, from which we obtained the stellar velocity dispersion map. \citep[see, e.g., ][for more details]{2022NewA...9701895L}

\begin{figure}
        \includegraphics[width=\columnwidth]{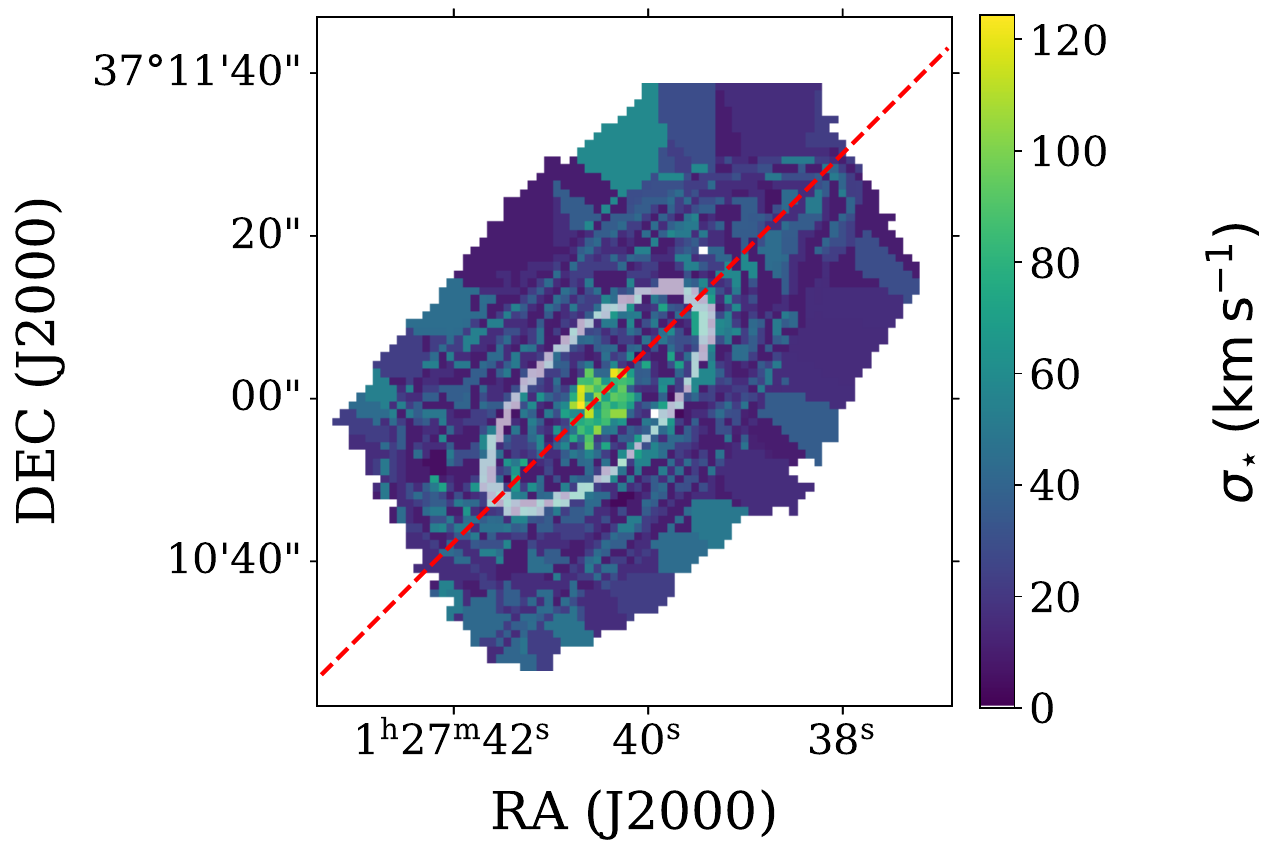}
    \caption{Stellar velocity dispersion map of NGC~551 from the third data release of the CALIFA survey. A velocity dispersion profile was generated by averaging velocity dispersion values in annular regions at different radial bins. The white elliptical region shows one such bin at 6 kpc. The dashed red line designates the major axis of the galaxy.}
    \label{fig:annuli_v}
\end{figure}

We used the second moment (MOM2) map to estimate the velocity dispersion in the stellar disk. We extracted a radial profile of the velocity dispersion by averaging $\sigma_\star$ values in elliptical annuli at different radii. These annuli were constructed considering the inclination and position angle of the stellar disk. In Fig.~\ref{fig:annuli_v}, we show one such annulus (white ellipse) at a radius of 6 kpc.

It is important to note that the measured MOM2 profile represents the intensity-weighted velocity dispersion along a line of sight. It is not the intrinsic velocity dispersion of the stellar disk. This profile can deviate significantly from the actual $\sigma_\star$ due to blending effects on the stellar spectra along a line of sight, particularly in the central regions where the rotation curve gradient is steep. Three primary factors contribute to this blending of the rotational velocity into the velocity width: the inclination of the stellar disk, the slope of the rotation curve, and the spatial resolution (PSF). These effects consistently result in overestimating the intrinsic $\sigma_\star$ profile. Hence, the velocity dispersion derived from the IFU second-moment map cannot be used directly in Eq.~\ref{eq:equation 5}. Instead, we used an iterative method (Sect.~\ref{sec:hydro}) to self-consistently assess the intrinsic $\sigma_\star$ while solving the hydrostatic equilibrium equation. Our estimation of $\sigma_\star$ was based solely on ensuring consistency between the hydrostatic solutions and the observations. However, we note that there are other methods by which the stellar vertical velocity dispersion can be estimated directly. For example, \cite{2010ApJ...716..234B} and \cite{2010ApJ...716....1B} estimated the z-component of $\sigma_\star$ by decomposing the line-of-sight velocity dispersion into three components (radial, azimuthal, and vertical). For NGC~551, our estimate is consistent with these studies (see Sect.~\ref{sec:vd_comp} for more details).

The determination of the vertical velocity dispersion in the atomic and molecular disks is relatively straightforward, as resolved spectroscopic observations of galaxies are possible using radio interferometers. Initial studies of \HI~emission in galaxy disks have indicated that the velocity dispersion ranges from $6 - 13$ \kms \citep{1984A&A...132...20S,1984A&A...134..258V,1993A&A...273L..31K}. A more recent high-spatial-resolution study of 12 nearby galaxies reported a consistent value of $\sigma_{\rm HI} = 11.9 \pm 3.1$ \kms \citep{2013AJ....146..150C}. Additionally, \citet{2016AJ....151...15M} found a similar result using stacking analysis, with $\sigma_{\rm HI} = 11.7 \pm 2.3$ \kms, further supporting the earlier findings.

Several studies have focused on estimating the velocity dispersion in the molecular disks in the Milky Way and external galaxies. For example, \citet{1984ApJ...281..624S} studied the velocity dispersion in molecular clouds of different masses in our Galaxy. They found that clouds with masses $10^2\:\mathrm{M}_\odot \leq M\:\leq 10^4\:\mathrm{M}_\odot$ have a velocity dispersion of $\sim 9.0^{+1.0}_{-1.1}$ \kms, and more massive clouds, having masses $10^4\:\mathrm{M}_\odot \leq M\:\leq 10^{5.5}\:\mathrm{M}_\odot$, have velocity dispersion of $\sim 6.6^{+0.9}_{-0.6}$ \kms. However, in a recent study done by \citet{2017A&A...607A.106M}, the molecular velocity dispersion inside the solar circle of the Galaxy was found to be $\sim 4.4 \pm\:1.2$ \kms. It should be noted that these studies measured velocity dispersion along directions primarily aligned with the midplane of the Milky Way. Hence, these values can differ from the actual vertical velocity dispersion.

\citet{1997A&A...326..554C} studied two nearby large spiral galaxies and found the molecular vertical velocity dispersion to be $\sim 6$ \kms~and $\sim 8.5$ \kms, respectively. They found this velocity dispersion to be nearly constant over the entire galaxy. Several other studies have also focused on nearby galaxies with high spatial and spectral resolution. For example, \citet{2011MNRAS.410.1409W} looked into the velocity dispersion of the molecular disks in 12 nearby spiral galaxies. They found an average molecular velocity dispersion of $6.1 \pm 1.0$ \kms~and additionally compared these velocity dispersion values with those in the atomic disks of the same galaxies. They found that the velocity dispersion in the atomic disk is almost two times higher than the average velocity dispersion in the molecular disk. Given these estimates, we adopt values of 6 and 12 \kms~for velocity dispersion in the molecular and atomic disks, respectively. 

We note that our choice of these values only marginally affects the calculated vertical density distribution in the stellar disk, similar to what is seen in \cite{2011MNRAS.415..687B}, where changes in the stellar velocity dispersion have minimal effect on the scale height of the atomic disk. \cite{2019MNRAS.484...81P} also showed similar effects on the atomic disk, where changes in the molecular velocity dispersion modify the atomic disk marginally. Similarly, it is to be expected that changes of the vertical velocity dispersion of the gas disks have a negligible effect on the stellar disk. For NGC~551, we varied the atomic and molecular velocity dispersion values within the range allowed by the literature ($4 - 8$ \kms~for the molecular disk and $10-14$ \kms~for the atomic disk). We find a less than 1\% change in the scale height of the stellar disk due to this variation.

\subsection{Rotation curve}
  \label{subsec:rotc}
 \begin{figure}
        \includegraphics[width=\columnwidth]{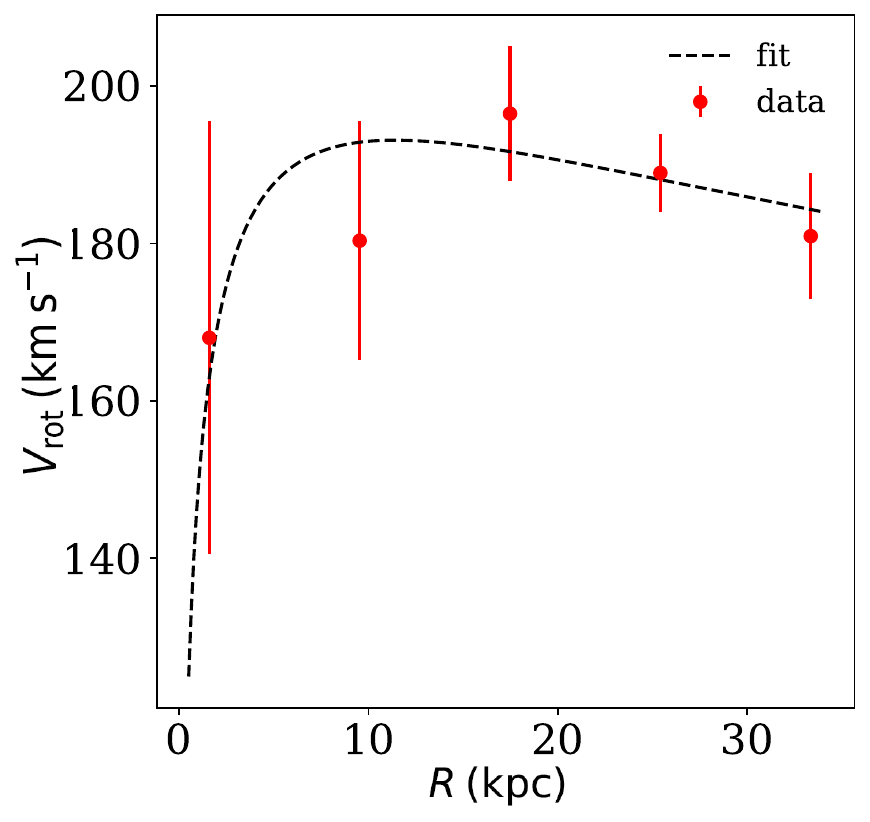}
    \caption{Rotation curve of NGC 551 derived from the \HI~spectral cube. The FAT pipeline was used to produce the rotation curve. The red circles with error bars represent the rotation curve, whereas the dashed black line signifies a fit to the rotation curve with a Brandt profile.}
    \label{fig:RC}
\end{figure}

\begin{figure*}
        \includegraphics[width=\textwidth]{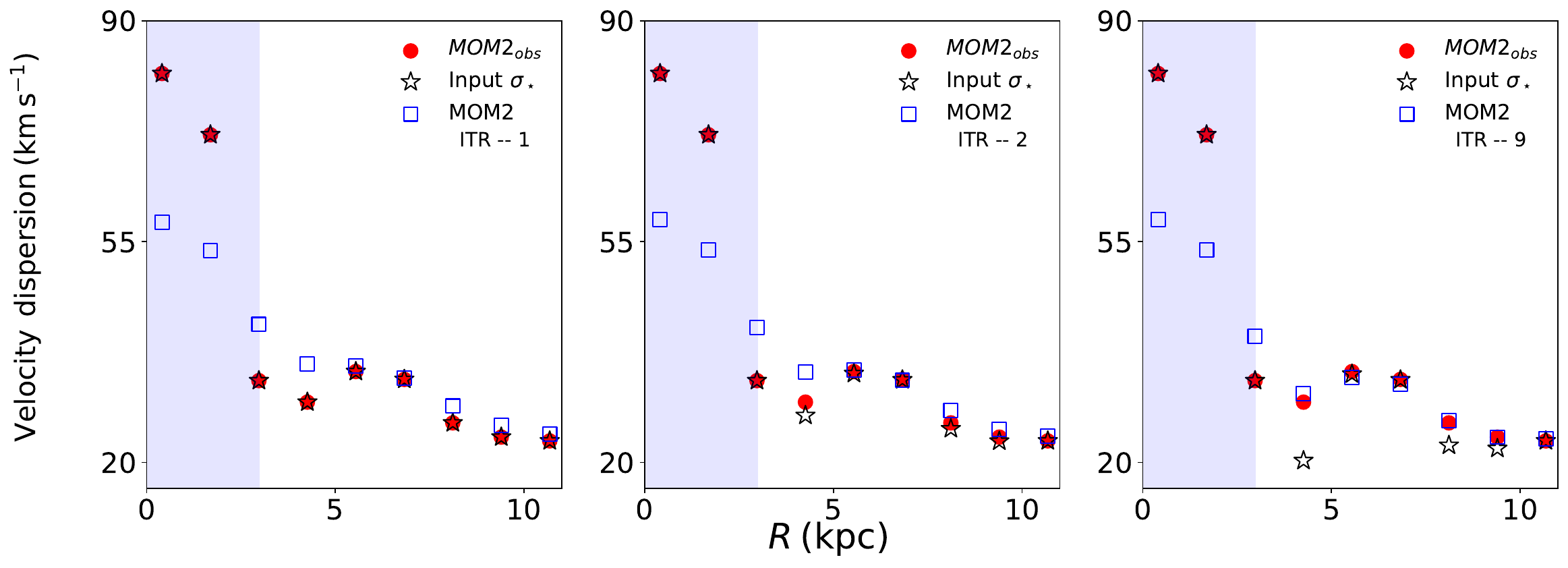}
    \caption{Illustration of the iterative method workflow. In each panel, the red circles represent the observed MOM2; the blue squares denote the simulated MOM2; and the black asterisks indicate the input intrinsic stellar velocity dispersion, $\sigma_\star$. The shaded region represents radii to which we do not apply iterative method. The left panel shows the results for the first iteration, whereas the middle and the right panel show the results for second and ninth iteration. The iterative method for NGC~551 quickly converged within ten iterations. See the text for more details.}
    \label{fig:itr}
\end{figure*}

The rotation curve is one of the required parameters in Eq.~\ref{eq:equation 5}. The rotation curve for NGC~551 was derived through tilted ring fitting of the \HI~spectral cube using TiRiFiC. The tilted ring model fitting is a widely used and standard method for performing kinematic modeling of disk galaxies \citep[see, e.g., ][]{1974ApJ...193..309R,2015MNRAS.451.3021D,2022MNRAS.513..168B}. In this method, the galaxy is assumed to be made up of concentric tilted rings, each with its own surface brightness, systemic velocity, position angle, and inclination. The parameters of these rings were fit against the observed spectral cube. The details of fitting the tilted ring and hence the kinematic modeling are described in Appendix \ref{sec:kin_mod}. In Fig.~\ref{fig:RC} we plot the rotation curve. In Eq.~\ref{eq:equation 5} we used the first derivative of the rotation curve. Considering the limitations in measurement, some sharp changes in the rotation curve can lead to an unrealistic value for its first derivative. To address this, we fit the rotation curve using a Brandt profile \citep{1960ApJ...131..293B} as

\begin{equation}
    V_{\rm rot} (R) = \frac{V_{\rm max} (R/R_{\rm max})}{\biggl(1/3 +2/3 \biggl( \frac{R}{R_{\rm max}}\biggr)^n \biggr)^{\frac{3}{2n}}},
\end{equation}

\noindent where $V_{\rm rot}$ is the observed rotation velocity, $V_{\rm max}$ represents the maximum achieved velocity, and $R_{\rm max}$ corresponds to the radius at which $V_{\rm max}$ is achieved. The parameter $n$ determines the rate at which the rotation curve rises to reach $V_{\rm max}$. The fit (dashed line) is shown in Fig \ref{fig:RC}. For NGC~551, we obtain $V_{\rm max} = 193.1 \pm 3.8$ \kms, $R_{\rm max} = 11.4 \pm 2.6$ kpc, and $ n = 0.25 \pm 0.09$. Adopting these parameters of the Brandt profile, we compute the derivative to estimate the radial term in Eq.~\ref{eq:equation 5}.

\subsection{Dark matter halo profile}
\label{sec:dm}
In Eq.~\ref{eq:equation 5} the dark matter halo serves as a crucial parameter that considerably contributes to the total gravity in the system, particularly at larger heights. We used the rotation curve of NGC~551, the stellar, \HI, and molecular surface density profiles to build the mass model and estimate the dark matter halo parameters.

Adopting an approach similar to the one described in \cite{2023MNRAS.524.6213B}, we used the stellar surface density profile to calculate the stellar circular velocity (circular velocity purely due to the gravity of the stellar disk) through Jeans anisotropic modeling (JAM) \citep{2020MNRAS.494.4819C}. In this analysis, we assumed a constant mass-to-light ({\it M/L}) ratio, which is treated as a free parameter in the JAM python package. The gas circular velocity from the gas surface density was calculated analytically assuming a thin disk. We then fit the total observed velocity, $V_{\rm rot}$, as a combination of the stellar, gas, and dark matter halo velocities components, following

\begin{equation}
    V_{\rm rot}^2 = V_{\rm \star}^2 + V_{\rm HI}^2 + V_{\rm H2}^2 + V_{\rm dm}^2,
\label{vel}
\end{equation}

\noindent through a parametric fit of the dark matter halo. 

We fit the total velocity $V_{\rm rot}$ by using a Markov Chain Monte Carlo (MCMC) sampler \citep{fox1998data} for doing mass modeling. To implement the MCMC fitting procedure, we utilized the code developed by \citet{TN} based on the MCMC method of \citet{2017MNRAS.464.1903K}. The mass modeling of NGC~551 uses the same procedure as described in \citet{2023MNRAS.524.6213B}. The figure of the modeled rotation velocity along with the velocity of other components can be found in Appendix \ref{sec:mass_mod_app}.

Many previous studies have shown that a Navarro-Frenk-White (NFW \cite{1997ApJ...490..493N}) profile can provide a good fit to the rotation curves of massive spiral galaxies, especially when baryonic effects are properly included \citep{2002ApJ...573..597K,2023MNRAS.524.6213B}. Although in some datasets, cored profiles (e.g., pseudo-ISO) performed equally well or marginally better \citep{2003MNRAS.339..243J,2008AJ....136.2648D}, the NFW profile was originally identified as the universal, spherically averaged density distribution emerging from $\Lambda$CDM N-body simulations \citep{1997ApJ...490..493N} and, thus, remains a widely used benchmark for halo modeling. Therefore, for NGC~551, we adopted an NFW profile and used MCMC optimization to determine the best-fit parameters for the dark matter component. The NFW density profile is given by the following equation:

\begin{equation}
    \rho_{\rm dm} (r) = \frac{\rho_0}{\frac{r}{R_{\rm s}}\Bigl(1+\frac{r}
    {R_{\rm s}}\Bigr)^2}.
    \label{eq:equation 8}
\end{equation}

\noindent For NGC~551, we obtained the dark matter halo parameters as $\rm \rho_0\:(\text{core density}) = 19.7 ^{+30.1}_{-24.1} \times 10^{-3} \ M_{\odot} \thinspace pc^{-3}$, and $R_{\rm s}\:(\text{scale radius}) = 11.4 ^{+6.2}_{-4.9}$ kpc. We used these parameters of the dark matter halo to solve Eq.~\ref{eq:equation 5} and estimated the 3D distribution of stars in NGC~551.

\section{Solving the hydrostatic equation}
\label{sec:hydro}

Equation \ref{eq:equation 5} represents three coupled second-order partial differential equations. We iteratively solved these equations by using a similar approach as adopted by many previous authors \citep{2002A&A...394...89N,2011MNRAS.415..687B,2020MNRAS.499.2063P}. As these equations are second-order, we required at least two boundary conditions to solve them, which can be given as 

\begin{align}
    (\rho_i)_{z=0} \:=  &\: \rho_{i,0}  && {\rm and} &\left( \frac{d\rho_i}{dz} \right)_{z=0} = &\: 0.
\label{eq:equation 11}
\end{align}

The density at the midplane of a galaxy was expected to be maximum due to vertical symmetry, satisfying the second condition. However, the midplane density ($(\rho_i)_{z=0}$) required for the first condition is not directly measurable in external galaxies. An iterative approach is employed to estimate it, leveraging the information of the observed surface density. The iterative method starts with an arbitrary trial midplane density at a particular radius and generates the solution. Then, the corresponding trial surface density is calculated by integrating this solution, $\Sigma_\star = 2 * \int_{0}^{\infty} \rho_{i,t} (z) \,dz$. By comparing this trial surface density with the observed one (at that radius), the midplane density is iteratively adjusted until the trial surface density matches the observation with high accuracy, typically within 0.1\%. For NGC~551, this iterative method converged within hundreds of iterations.

As discussed in Sect.~\ref{sec:VD}, the observed stellar velocity dispersion, $\sigma_\star$, is the MOM2 of the line-of-sight spectra (MOM2). This MOM2 is an overestimate of the intrinsic velocity dispersion. Hence, using it directly in Eq.~\ref{eq:equation 5} would produce an incorrect 3D density of the stars in the stellar disk. To address this issue, we used an iterative approach to correctly guess the intrinsic $\sigma_\star$, such that it produces the observed MOM2 when modeled self-consistently. We used the same approach as used by \citet{2020MNRAS.499.2063P} \citep[see also][]{2021MNRAS.501.3527P} to iteratively solve for the densities as well as for the intrinsic velocity dispersion in NGC~551.

In the first iteration, we began with the observed MOM2 profile as the intrinsic velocity dispersion and solve Eq.~\ref{eq:equation 5} between radii $ \mathrm{2} \leq R \leq \mathrm{11}$ kpc to generate the stellar densities as a function of radius ($R$) and height ($z$). We built a 3D dynamical model of the stellar disk with this density distribution and the observed rotation curve. Next, we inclined this 3D model to the observed inclination ($63\degree$), project it into the sky plane, and generate a spectral cube. This spectral cube was then convolved with the observed PSF of $2.28\:$\arcs~$(\sim 796\: \text{pc})$, and collapsed to generate simulated moment maps. The simulated MOM2 map was then used to construct a MOM2 profile, which is equivalent to an observation. In the first iteration, as we began with the MOM2 profile, which is already an overestimate of the $\sigma_\star$, the simulated MOM2 would be higher than the observed one. In the next iteration, we adjusted the observed MOM2 values (points outside the shaded region) by scaling them with a factor determined by the difference between these values and the $\sigma_\star$ from the previous iteration. We note that this multiplicative factor is different at different radii, ensuring faster convergence.

In the subsequent iterations, as the input MOM2 profile kept decreasing, the simulated MOM2 approached the observed one. We stopped this method when the simulated and the observed MOM2 values match within 5\%. At this point, the input MOM2 can be considered the intrinsic $\sigma_\star$, producing observed MOM2 self-consistently. The left panel of Fig.~\ref{fig:itr} shows the first iteration, where the observed MOM2 is considered as the input $\sigma_\star$, resulting in a simulated MOM2 higher than the observed. In the subsequent iterations (middle and right panels), a lowered input $\sigma_\star$ results in simulated MOM2 profiles that are close to the observed one. We find that this method converged within 9 iterations (with $\sim 5\%$ accuracy). We emphasize that the MOM2 points are sampled at approximately 1.7 times the PSF, or about 3.9\arcs~apart, to prevent divergence related to beam smearing. When generating the final simulated spectral cube, the raw-resolution cube was convolved with the PSF. If the MOM2 points are sampled too closely, any variation in MOM2 values at one radius could affect neighboring points due to convolution, potentially causing the iterative method to fail. For NGC~551, we find that sampling at $\approx 1.7$ times the PSF avoids this divergence.

We note that in the central regions of galaxies, factors such as the steep gradient in the rotation curve, strong noncircular motions, and the presence of a bulge can significantly broaden the stellar spectra. This artificial broadening can dominate the spectral width, particularly in galaxies with high inclinations, compared to the actual value of $\sigma_\star$. Consequently, the effective MOM2 can become insensitive to the intrinsic $\sigma_\star$, causing the iterative method to diverge. To avoid these complications, we exclude the central region of NGC~551 from our modeling. We did not solve Eq.~\ref{eq:equation 5} at radii $R < 2$ kpc. Moreover, the molecular disk in NGC~551 was extended to $R \sim 11\,\text{kpc}$. Hence, we solved Eq.~\ref{eq:equation 5} between radii $ \mathrm{2} \leq R \leq \mathrm{11}$ kpc, at every 100 pc. The shaded region in Fig.~\ref{fig:itr} represents the radii where we did not apply the iterative method. The spatial resolution of the IFU data is $\sim 796$ pc at the galaxy distance of 72 Mpc, based on the $2.28\:$\arcs~full width at half maxima (FWHM) of the PSF as recorded in the pyPIPE3D FITS header. Hence, solving Eq.~\ref{eq:equation 5} every 100 pc is adequate in building the dynamical model. However, in the vertical direction, we actually used an adaptive resolution to capture the variation of the density at a small height (within a kpc). At every iteration, the step size was determined by the slope of the density solutions in the previous iteration \citep[see, e.g., ][for more details]{2018MNRAS.478.4931P}. For NGC~551, we find that the vertical resolution is always better than 5 pc.

\section{Results and discussion}

\begin{figure}
        \includegraphics[width=\columnwidth]{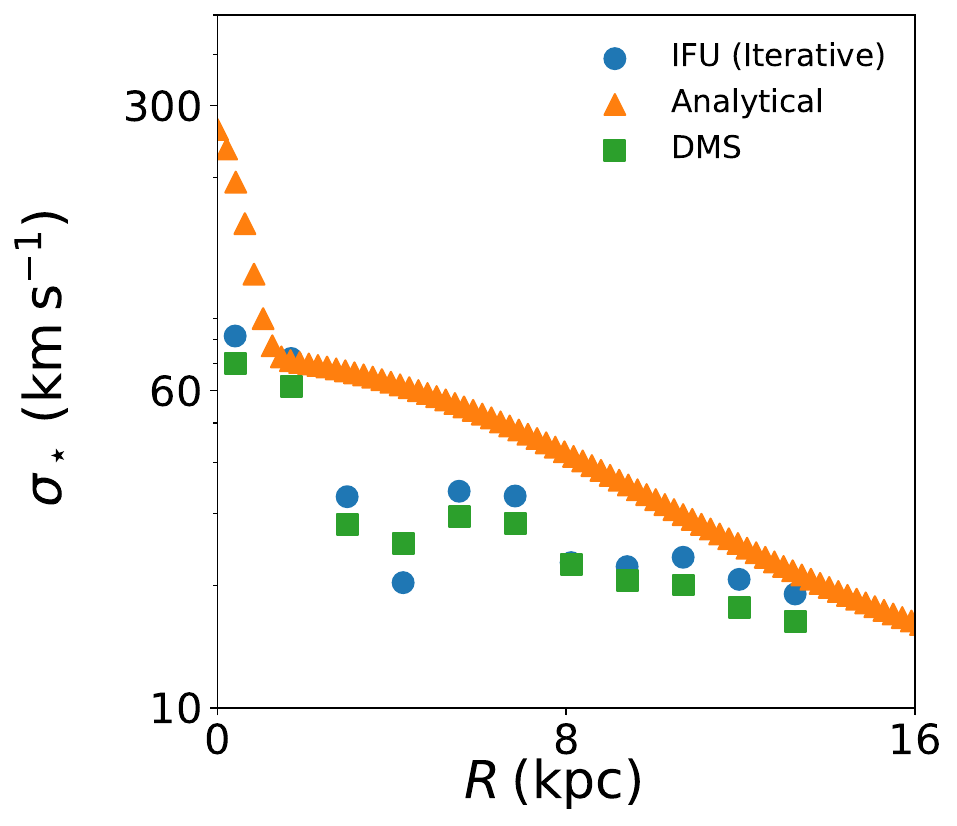}
    \caption{Velocity dispersion profiles of the stellar disk in NGC~551. The orange triangles represent the $\sigma_\star$, as obtained by analytical calculation from \citet{2008AJ....136.2782L}. The blue circles show the intrinsic $\sigma_\star$ profile, as obtained by the iterative method. The green squares show the $\sigma_\star$ profile obtained with the DMS formalism \citep{2010ApJ...716..234B}. The analytical expression can significantly overestimate $\sigma_\star$ at all radii.}
    \label{fig:sigma_comp}
\end{figure}

\begin{figure*}
        \includegraphics[width=\textwidth]{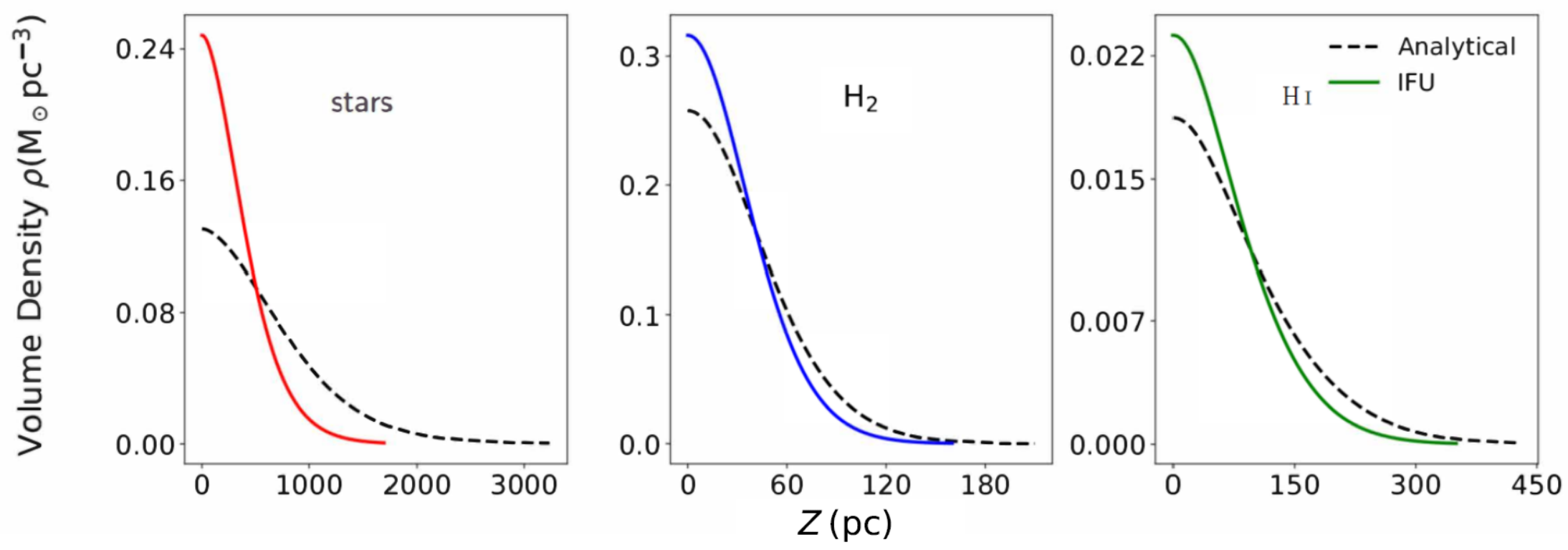}
    \caption{Solutions to the hydrostatic equilibrium equation at a radius of 6 kpc. The left, middle, and right panels present density solutions for stars, molecular, and atomic gas, respectively. The solid lines in each panel represent solutions when the hydrostatic equations were solved with the iterative method using CALIFA IFU data. The dashed lines represent the solutions with the analytically calculated stellar velocity dispersion described by Eq.~\ref{eq:analytical_expressiom}. The choice of stellar velocity dispersion greatly influences the density solutions in all three baryonic disks.}
    \label{fig:vol den1}
\end{figure*}

\subsection{The stellar velocity dispersion profile}
\label{sec:vd_comp}

As discussed in Sect.~\ref{sec:VD}, spectroscopic observations enable us to measure the intrinsic velocity dispersion, even though the observed values are blended by line-of-sight projection effects. In Fig.~\ref{fig:sigma_comp}, we plot the stellar velocity dispersion in NGC~551 as obtained by the iterative method (blue circles), along with the analytically calculated dispersion using the Eq.~\ref{eq:analytical_expressiom}. A scale length value of $4.6$ kpc was used to obtain the analytical $\sigma_{\rm \star}$ (\cite{2008AJ....136.2782L} orange triangles). This scale length was derived from a 1D fit to the $\Sigma_\star$ profile obtained from MGEfit (for further details and plots, see Appendix \ref{sec:scale_length}). As shown, for NGC~551, the intrinsic $\sigma_\star$ can differ from the analytical value by as much as 70\%. This, in turn, can change the stellar scale height by a factor of $1.5-2.0$ (see, e.g., Fig.~\ref{fig:scale height1}). Hence, the analytical estimation of the vertical velocity dispersion in the stellar disks might often overestimate the thickness of the disks.

In the literature, other methods have also been used to estimate the stellar velocity dispersion. For example, the Disk Mass Survey (DMS) \citep{2010ApJ...716..198B} observed 46 face-on galaxies with the CALIFA IFU. The line-of-sight stellar velocity dispersion was measured as the projection of the stellar velocity ellipsoid (SVE) along the line of sight. By applying the principal SVE ratios, a relation was derived between the projected and deprojected velocity dispersions \citep[see Eqs. 2, 3, and 4 in Sect.~3.5.4 of][for more details]{2010ApJ...716..234B}. Using the vertical anisotropy ($\beta_z$ = $0.388 ^{+0.007}_{-0.023}$ from \cite{2017MNRAS.469.2539K} and the median of the azimuthal anisotropy ($\beta_\phi$ profile = $0.2 \pm 0.03$ from \citealt{2018A&A...618A.121C}), the principal SVE ratios ($\frac{\sigma_z}{\sigma_R}$ $\&$ $\frac{\sigma_\phi}{\sigma_R}$) were estimated. These $\frac{\sigma_z}{\sigma_R}$ and $\frac{\sigma_\phi}{\sigma_R}$ values were used in Eq.~5 of \cite{2010ApJ...716..234B}, to calculate the stellar velocity dispersion in NGC~551. In Fig. \ref{fig:sigma_comp}, the green squares represent $\sigma_\star$ as obtained by the DMS formalism. For NGC~551, the DMS $\sigma_\star$ is largely consistent with the iterative method within $\sim 20-25\%$. However, it should be noted that the DMS sample consists primarily of face-on galaxies, and their method is best suited for low-inclination galaxies in the $25\degree$ to $35\degree$ range, whereas NGC~551 has an inclination of $63\degree$.

\subsection{3D density distribution in the stellar disk}

\begin{figure*}
        \includegraphics[width=\textwidth]{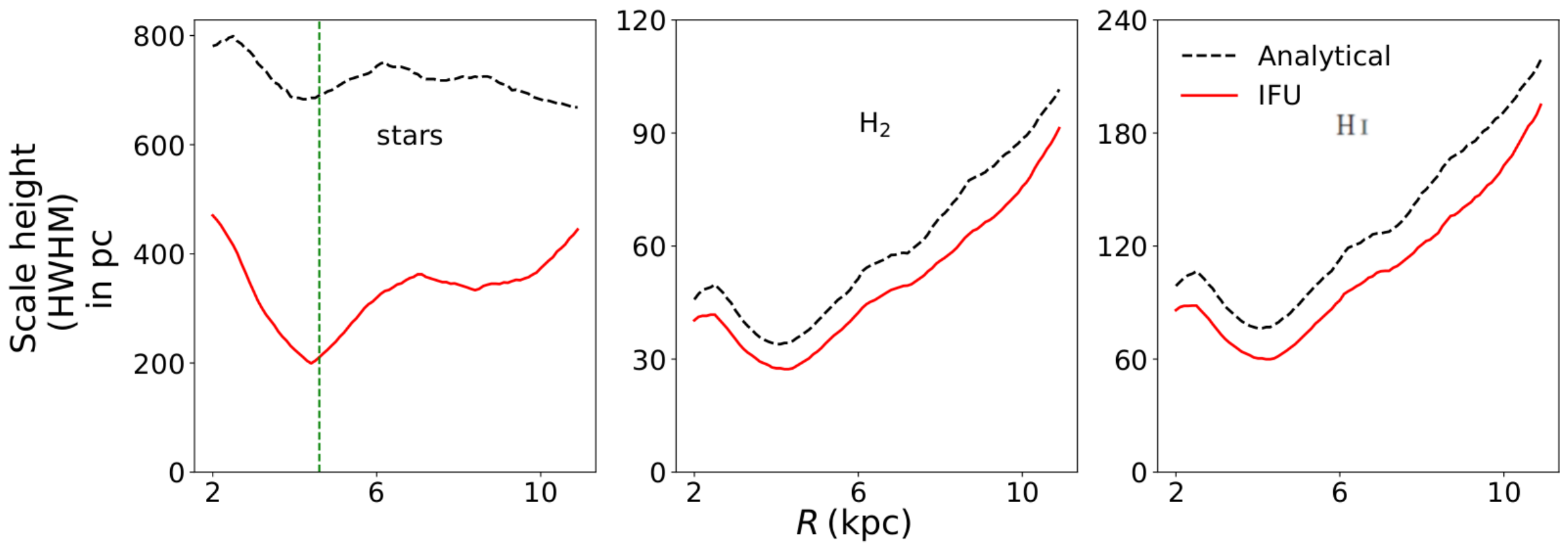}
    \caption{Scale height of different baryonic disks in NGC~551 under hydrostatic equilibrium. The left, middle, and right panels represent scale heights for the stellar, molecular, and atomic disks, respectively. The solid lines in each panel represent the scale height due to our iterative method, whereas the dashed lines show the scale heights for the analytically calculated $\sigma_\star$. The vertical dashed line in the left panel indicates the scale length of the stellar disk (4.6 kpc). The analytically calculated $\sigma_\star$ overestimates the stellar scale height by more than a factor of two, giving the false impression of a thick stellar disk.}
    \label{fig:scale height1}
\end{figure*}

The solutions of the hydrostatic equation provide the volume densities of different disk components as a function of $R$ and $z$. We solve this equation every 100 pc from a radius of 2 kpc to 11 kpc. Fig.~\ref{fig:vol den1} shows the density solutions (solid lines) for the different disks in NGC~551 at a radius of 6 kpc. We also compare these solutions with the stellar velocity dispersion derived from the analytical expression obtained from Eq.~\ref{eq:analytical_expressiom} (dashed lines). As shown in the figure, the analytically calculated $\sigma_\star$ produces lower densities at the midplane, and consequently, the disk extends to larger heights. This produces thicker stellar disks than expected. Moreover, as the other disks (atomic and molecular) are coupled to the stellar disk through gravity, they also become thicker than expected. In NGC~551, at a radius of 6 kpc, the midplane densities of the stellar, atomic, and molecular disks obtained via numerical computation differ from the analytically derived values by approximately 91\%, 25\%, and 22\%, respectively.

The half-width at half maximum (HWHM) is an excellent measure of the thickness of a baryonic disk. This HWHM is often adopted as the scale height of the density distribution in the vertical direction. Using our solutions, we estimated the scale height of different disk components as a function of radius. In Fig.~\ref{fig:scale height1}, we plot the scale height of different disk components (solid lines) in NGC~551. We also compare the scale height profiles for solutions with analytically calculated $\sigma_\star$, shown by dashed lines. As shown in the figure, the scale height of the stellar disk can be overestimated by a factor of two when using an analytically calculated $\sigma_\star$. This difference is lower in the atomic and molecular disks; however, it can also make a difference of $\sim 25-30 \%$. This indicates that the determination of the stellar velocity dispersion by the spectroscopic method is crucial in assessing the vertical density distribution in the baryonic disks (especially in the stellar disk) in galaxies.

\begin{figure*}
        \includegraphics[width=\textwidth]{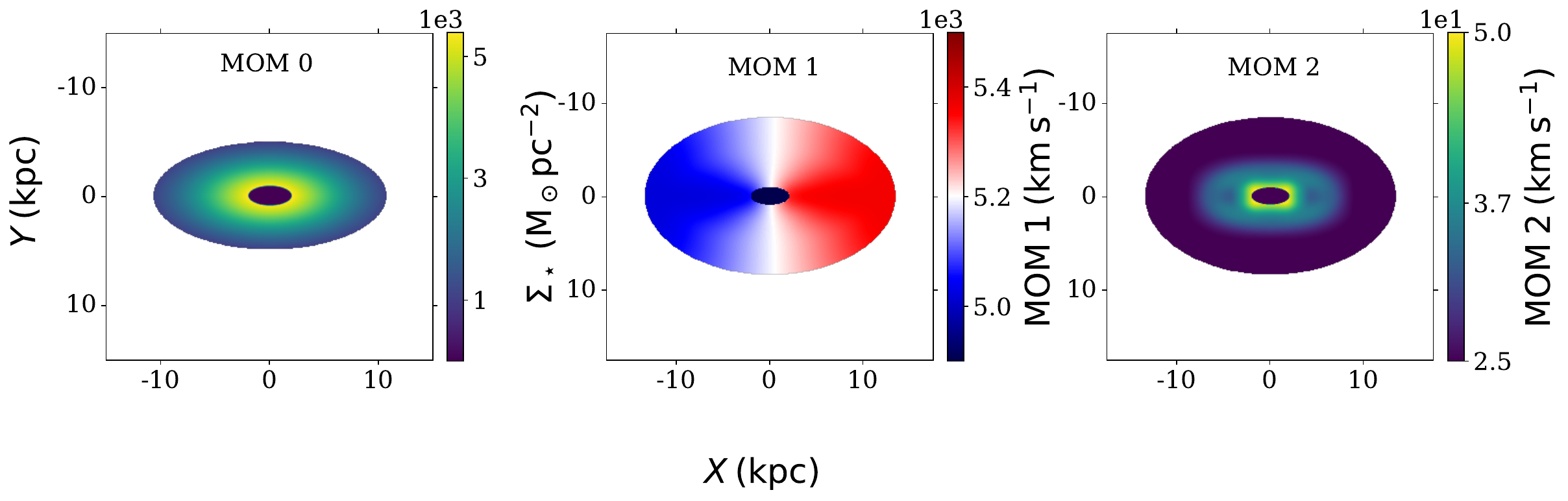}
    \caption{Simulated moment maps of the stellar disk in NGC~551. The left panel shows the moment zero map, i.e., the total surface density map in the units of $\rm M_\odot\:pc^{-2} $. The middle panel shows the moment one map or the velocity field of the galaxy. The right panel represents the velocity dispersion map. The last two maps have units of \kms. These moment maps were generated using the simulated spectral cube obtained through the dynamical modeling of the stellar disk.}
    \label{fig:mom cube1}
\end{figure*}

\begin{figure*}
        \includegraphics[width=\textwidth]{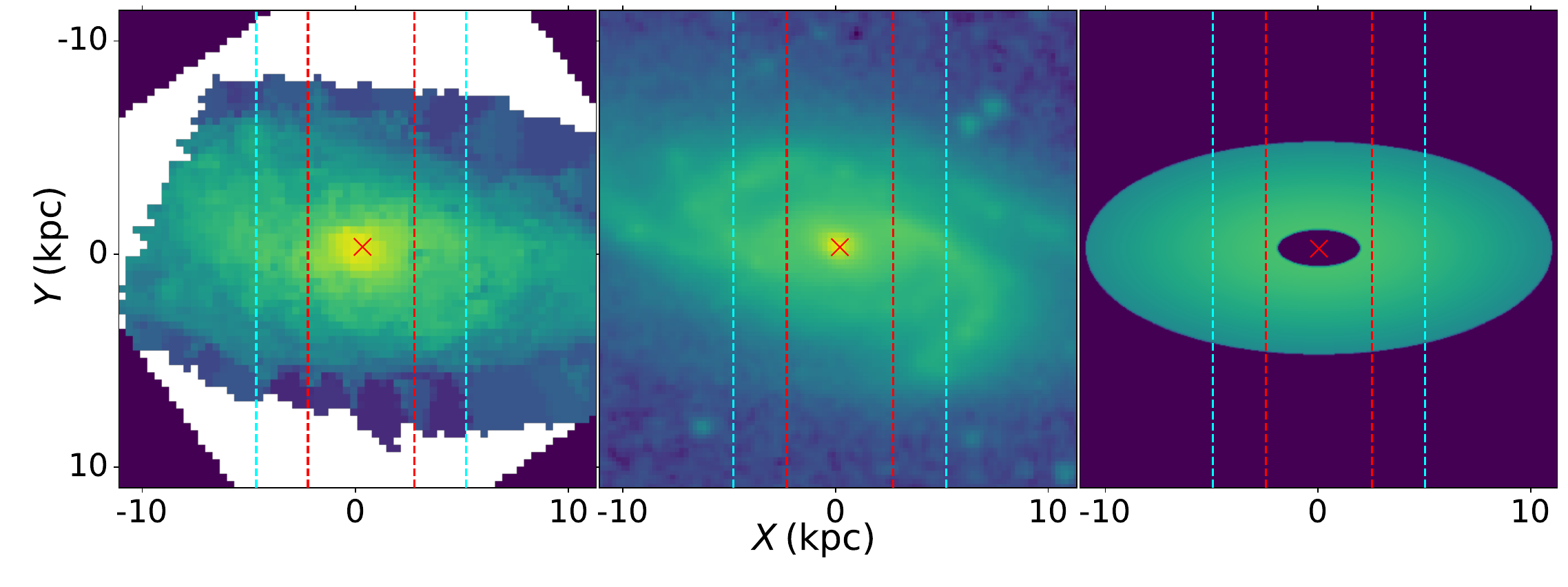}
    \caption{Cuts along the y-axis at radii of 2.5 and 5.0 kpc (dashed red and cyan lines) for comparison between the observations and the simulation. The left panel represents the surface brightness map of the stellar disk, as obtained by IFU observation. The middle panel shows the stellar disk as observed in \textit{Spitzer} 3.6\,$\mu{\rm m}$ data. The right panel indicates our simulated intensity distribution. The crosses in each panel indicate the center of the galaxy.}
    \label{fig:vc}
\end{figure*}

\begin{figure*}
        \includegraphics[width=\textwidth]{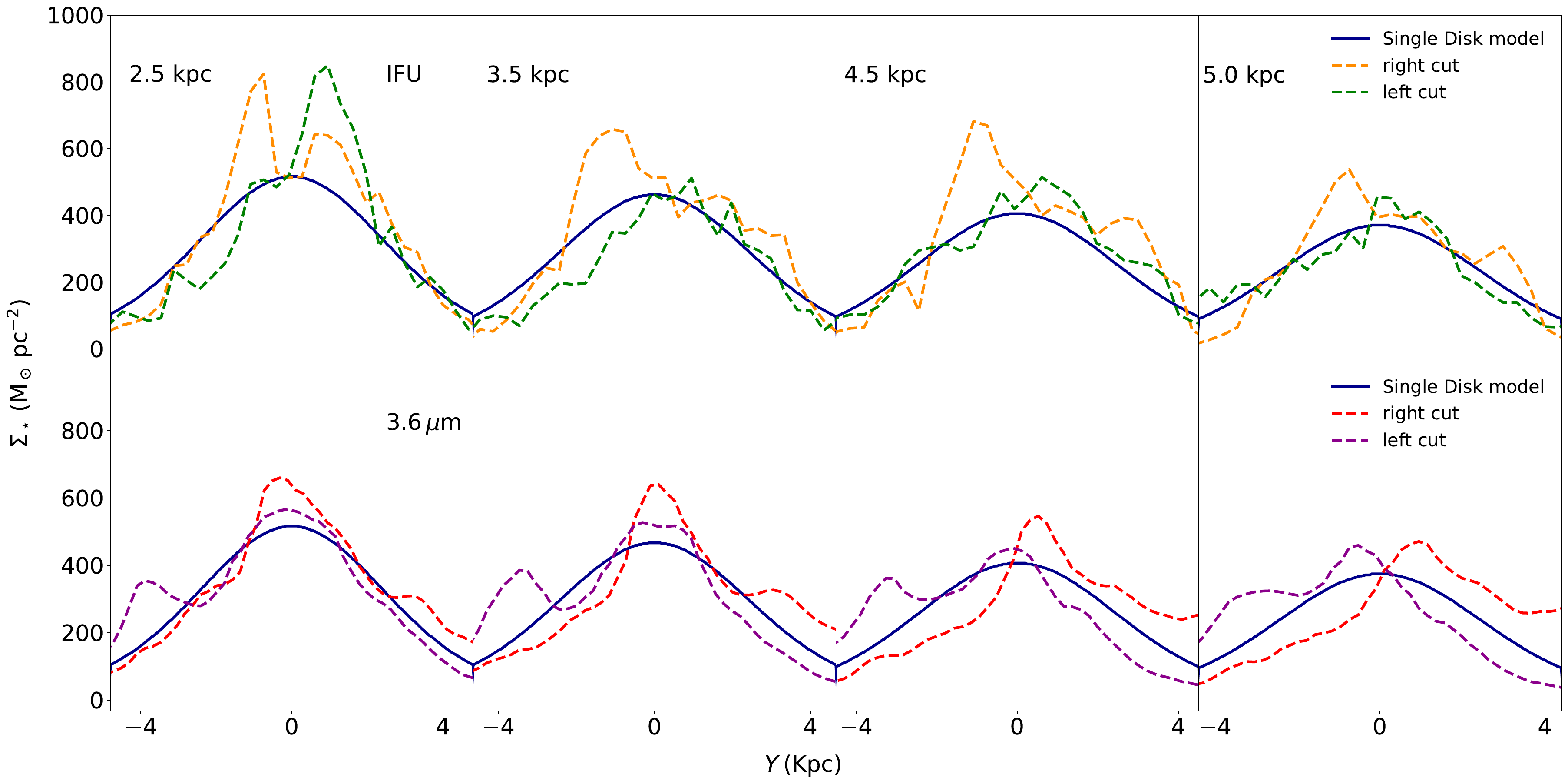}
    \caption{Comparison of cut profiles along the y-axis of the surface density distribution in the stellar disk of NGC~551. In the top panels, we compare the IFU observation with the simulation, whereas in the bottom panels, we compare \textit{Spitzer} 3.6\,$\mu{\rm m}$ with the simulation. Each column of panels shows different radii. In each panel, the solid lines denote the profiles for simulated maps, whereas the dashed lines show the observed ones. Two different colors represent two different halves (left and right) of the observed map as shown in Fig.~\ref{fig:vc}. The simulated profiles at different radii match the observation reasonably well. See the text for more details.}
    \label{fig:single_disk}
\end{figure*} 

\begin{figure}
        \includegraphics[width=\columnwidth]{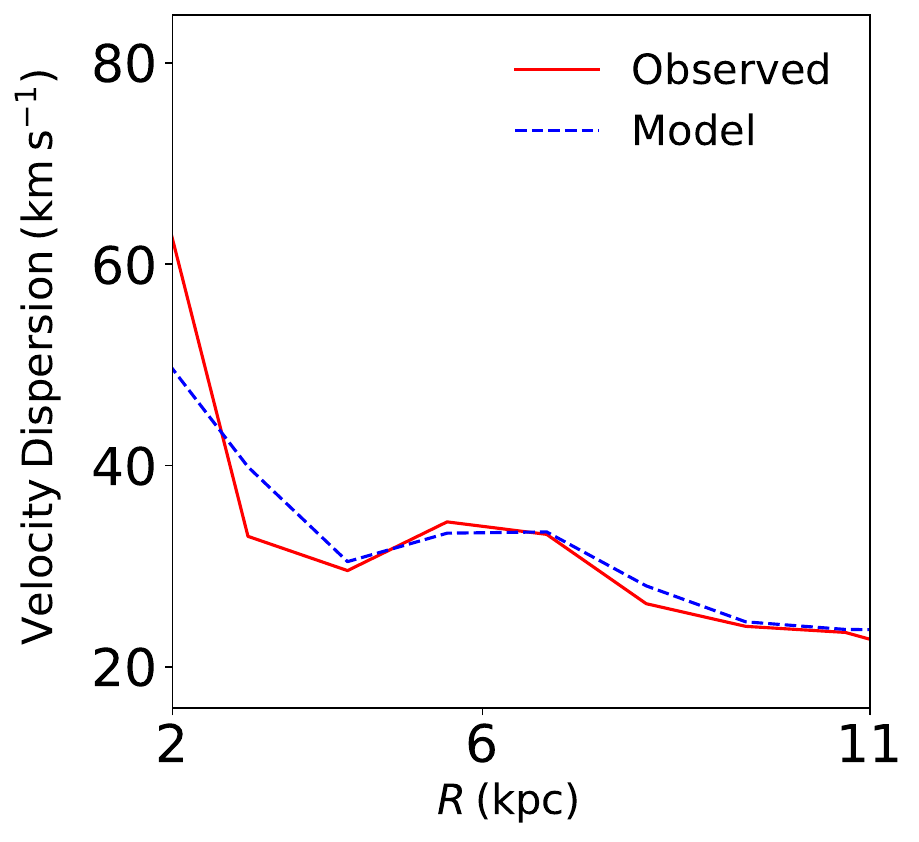}
    \caption{Radially averaged velocity dispersion profiles. The solid red line represents the observed MOM2 obtained from CALIFA IFU. The dashed blue line indicates the MOM2 obtained from the MOM2 map shown in Fig.~\ref{fig:mom cube1}.}
    \label{fig:vel_disp}
\end{figure}

\subsection{Dynamical model of the stellar disk}

Next, we used the density solutions of the stellar disk and the observed rotation curve to build a 3D dynamical model of NGC~551. We incline this dynamical model to the observed inclination of $63\degree$, project it to the sky plane, and construct a spectral cube. This cube is then convolved with the PSF of the IFU observation, and consequently, moment maps are made. These model moment maps are equivalent to an observation.

In Fig.~\ref{fig:mom cube1}, we show our model moment maps of the stellar disk in NGC~551. These maps capture the kinematics and the surface density distribution of the stellar disk. The MOM-0 map shows the stellar surface density distribution, the MOM-1 map shows the velocity field, while the MOM-2 map represents the line-of-sight weighted velocity dispersion field. We compare these simulated maps with real observations to check the consistency of our modeling. We obtain the observed stellar surface density map by processing IFU data using pyPipe3D. Furthermore, we obtain the stellar surface density map utilizing \textit{Spitzer} data (using the calibration given in \citealt{2012AJ....143..139E}). We take cuts along the y-axis on the model and the observed surface density maps (Fig.~\ref{fig:mom cube1}) at different radii and compare the resulting profiles. In Fig.~\ref{fig:vc}, we show two representative cuts at a radius of 2.5 kpc and 5.0 kpc. In Fig.~\ref{fig:single_disk}, we show the resulting surface brightness profiles at different radii. We take cuts at radii of 2.5, 3.5, 4.5, and 5 kpc. We compare our model (blue lines) with surface density profiles from IFU (broken lines in top panels) and the 3.6\,$\mu{\rm m}$ maps (broken lines in bottom panels). As shown in the figure, the model profiles match the observation reasonably well. However, we note that at the central region, the observed map has an extra emission, which is not shown in the modeled map. This emission is more prominent in the IFU map than in the 3.6\,$\mu{\rm m}$ map. Furthermore, both the top and bottom panels show that as the radius increases, the amplitude of this spiky component reduces.

The optical disk of NGC~551 has an inclination of 63$^\circ$. This non-edge-on orientation can imprint signatures of in-plane, unsmooth structures onto the vertical profiles. The departures observed in the central region are expected, primarily from non-axisymmetric structures such as spiral arms (visible in the middle panel of Fig.~\ref{fig:vc}), which our axisymmetric model does not account for.

The velocity dispersion, as traced by the MOM2, also serves as a critical quantity for inspecting the consistency of the dynamical modeling. In Fig.~\ref{fig:vel_disp}, we compare the MOM2 profiles derived from the modeled and observed maps. As shown, the profile obtained from the modeled MOM2 map (rightmost panel in Fig.~\ref{fig:mom cube1}) closely matches the observed one, derived using IFU data. This indicates that our modeling reproduces the dynamics of NGC~551 reasonably well.

\subsection{Edge-on model}

\begin{figure*}[!htp]
        \includegraphics[width=\textwidth]{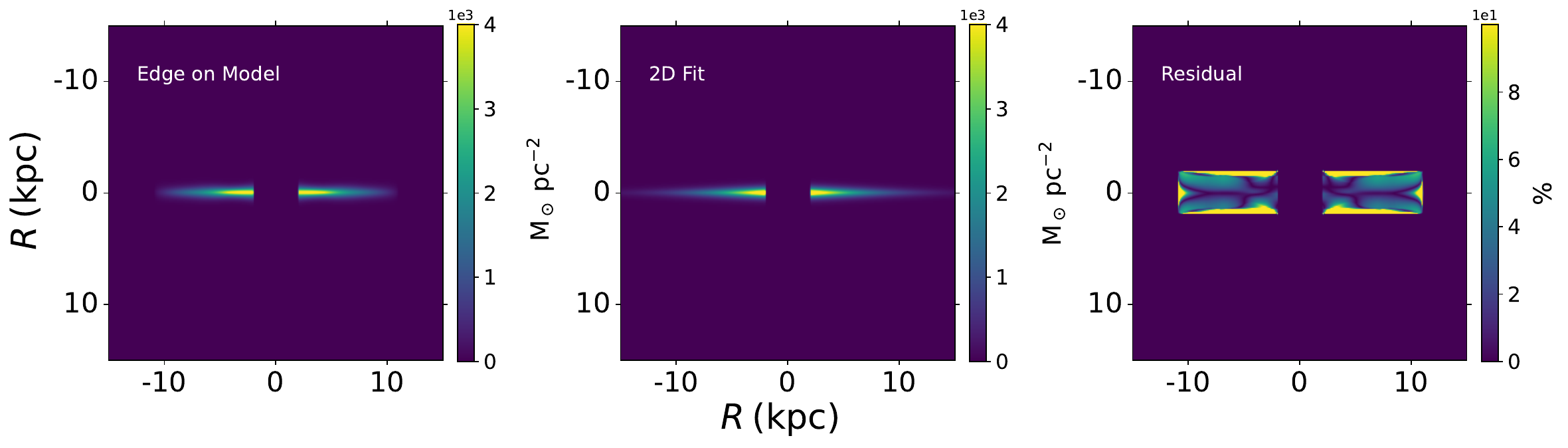}
    \caption{Surface brightness map of the stellar disk in NGC~551 in an edge-on orientation. The left panel shows the surface brightness map, whereas the middle panel represents a 2D fit to the map (using Eq. \ref{eq:2d_fit_sech}). The right panel shows the residual of the fitting. The masked portion in the middle of each panel represents the region where we did not solve the hydrostatic equilibrium equation. The 2D fit describes the simulated data reasonably well.}
    \label{fig:2dfit}
\end{figure*}

Several studies have focused on the stellar surface density distribution of edge-on galaxies to investigate multicomponent stellar disks \citep{2006AJ....131..226Y,2011ApJ...741...28C} in external galaxies. These studies find that a two-component stellar disk better describes the observed stellar surface density. However, the observed surface density for an edge-on disk is the sum of all the light along a sight line from different radii. Furthermore, as the scale height of the stellar disk changes as a function of radii (for both the thin and the thick disk), the integrated light might not reflect the true nature of the disk.

To test this, we incline our 3D dynamical model to an angle of $90\degree$ and project it to the sky plane to produce an edge-on surface brightness map. This is equivalent to how we would observe NGC~551 in an edge-on orientation. We then used this map as a template to investigate whether traditional methods of fitting multicomponent stellar disks to the edge-on surface brightness distribution can recover the original disk parameters.

To do so, we adopted a widely used approach and fit the galaxy surface brightness with a 2D edge-on model for surface brightness given by  
\begin{equation}
    \Sigma(R, z) = \Sigma_0\:(R/h_{\rm R})\:K_1(R/h_{\rm R})\:\operatorname{sech^2} (z/h_{\rm Z}) ,
\label{eq:2d_fit_sech}
\end{equation}
where $\Sigma_0$ is the edge-on peak surface brightness, $h_{\rm R}$ and $h_{\rm Z}$ are the scale length and the scale height, respectively, and $K_1(x)$ is the modified Bessel function of the first kind \citep{1981A&A....95..105V,2006AJ....131..226Y,2011ApJ...729...18C,2020MNRAS.494.1751M}. This method estimates the scale length and the scale height of the observed disk. In Fig.~\ref{fig:2dfit}, we show the model surface brightness distribution of NGC~551 in edge-on orientation in the left panel and a 2D fit to it in the middle panel. We also show the residual in the figure on the right panel. As shown, the edge-on surface brightness is well fit by the 2D function. From the fit, we find the scale length of the stellar disk to be $\sim 3.2$ kpc.

In contrast, we obtain a scale length of $\sim 4.6$ kpc from a 1D fit to the stellar surface density profile derived via MGE fit (blue plot in Fig. \ref{fig:SD}; for further details, see Appendix \ref{sec:scale_length}). Notably, \cite{2004A&A...414..905H} performed photometric analysis on $H-\text{band}$ data to obtain a similar scale length of $\sim 4.5$ kpc, in agreement with our findings. This indicates that at edge-on orientation, due to the line-of-sight integration effect, a 2D fit does not recover the true scale length of the disk. From the 2D fit, the scale height of the stellar disk was found to be $\sim 400$ pc, which reasonably matches the actual scale height, which varies between $200 - 470$ pc as shown in Fig.~\ref{fig:scale height1}.

\begin{figure}
        \includegraphics[width=\columnwidth]{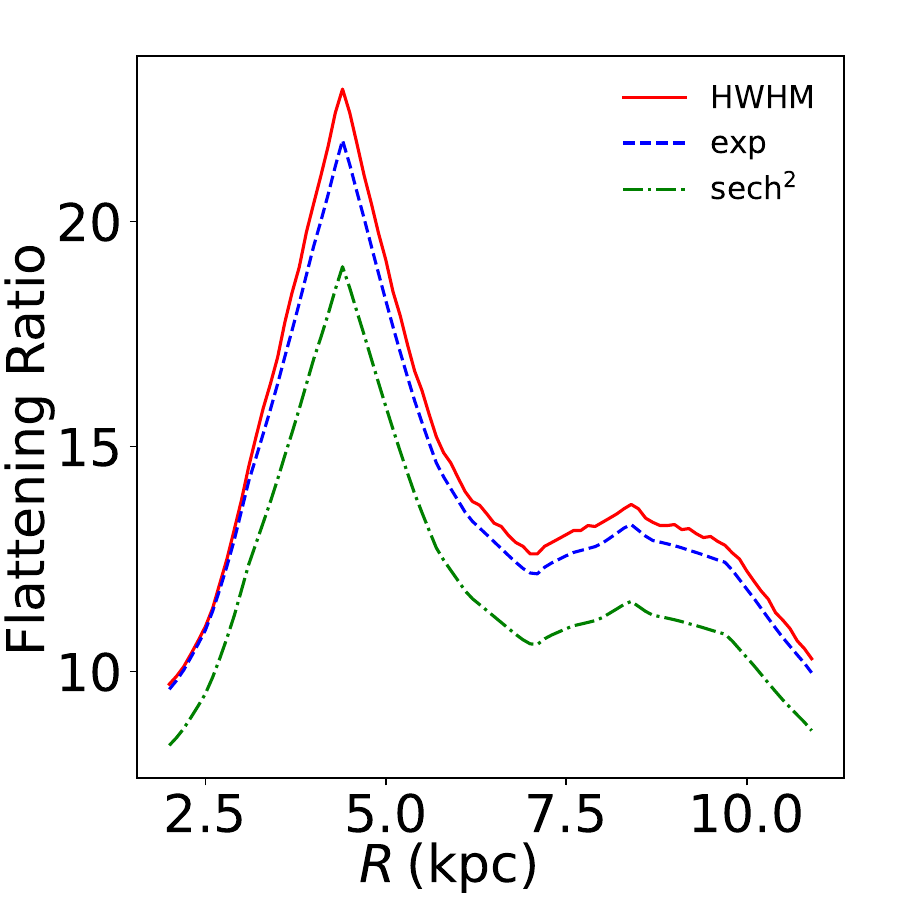}
    \caption{Estimated FR of the stellar disk in NGC~551 as a function of radius. The estimation was obtained using different modeled scale heights. The solid red line represents the FR of scale length and HWHM scale height. The dashed blue line represents the FR obtained from the ratio of scale length and exponential scale height. The dashed-dotted green line represents the FR obtained from the ratio of scale length and $\rm sech^2$ scale height. The FR varies substantially (between $\sim 8-23$) with radius.}
    \label{fig:fr}
\end{figure}

Although the recovered scale height matches the actual values reasonably, an underestimated scale length can significantly alter the computed FR, defined as scale-length divided by scale-height. We estimated the scale height as the HWHM of the vertical density distribution. However, in many studies, the scale height is defined as the decay constant of an exponential or $\rm sech^2$ fit to the vertical density profile. For completeness, we also estimated the FR by adopting these definitions of scale height. In Fig.~\ref{fig:fr}, we plot FR as a function of radius for NGC~551. As shown, for NGC~551, the FR values vary between 8 and 23 depending on the radius.

Several studies used an exponential function instead of a $\rm sech^2$ to describe the vertical brightness profile of galaxies and fit the edge-on surface brightness. For example, \citet{2002MNRAS.334..646K} fit the surface brightness of 34 edge-on galaxies with a 2D edge-on model for surface brightness given by
\begin{equation}
   \Sigma(R, z) = \Sigma_0\:(R/h_{\rm R})\:K_1(R/h_{\rm R})\:\operatorname e^{-(z/h_{\rm Z})}.
\label{eq:2d_fit_exp}
\end{equation}
They used the fit parameters to estimate the FR in these galaxies. They found an average FR value of $7.3 \pm 2.2$. Furthermore, \citet{2014ApJ...787...24B} fit the SDSS $g-, r-, \text{and}\:\: i-\text{band}$ surface densities of 4768 edge-on galaxies with the same model described by Eq.~\ref{eq:2d_fit_sech} and estimated the FRs. In Fig. 13 of their paper, they plot a histogram of FR of a subsample of 3,865 edge-on galaxies. Their FR values range from as low as 2 to as high as 20, with $\sim 100$ galaxies having ${\rm FR} \sim 8$. For consistency, we also fit the edge-on surface density of NGC~551 with their method and found an FR value of $10.48 \pm 0.01$ for the model given by Eq.~\ref{eq:2d_fit_exp} and $7.88 \pm 0.02$ for the model given by Eq.~\ref{eq:2d_fit_sech}. These values are consistent and fall well within the range of their FR values. However, it should be noted that these values differ considerably from the actual FR calculated using scale height data. The FR, assuming an exponential definition of scale height, is found to be $\sim 9 - 22$, while for a $\rm sech^2$ definition, it is $\sim 8 - 19$. These results suggest that relying solely on a 2D fit to the edge-on surface density can lead to a systematic underestimation of the stellar disk's true FR, as it does not fully capture the complexities of the disk’s vertical structure.

\begin{figure*}[!htp]
        \includegraphics[width=\textwidth]{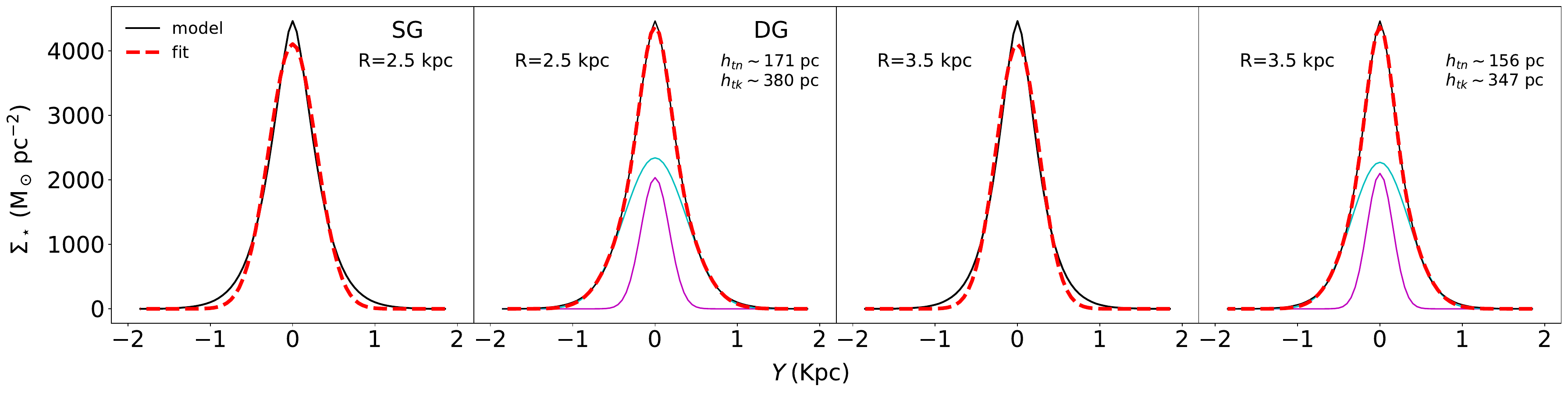}
    \caption{Vertical cuts of the simulated intensity distribution of the stellar disk in an edge-on orientation and its fit with single and double Gaussian components. The first two panels show vertical cuts at a radius of 2.5 kpc, and the last two panels show cuts at a radius of 3.5 kpc. The solid black lines represent the vertical cuts of the edge-on intensity profiles. The dashed red lines represent the fits. The first and third panels show a single Gaussian fit, whereas the second and fourth panels show a double Gaussian fit. The double Gaussian fits represent the data better, despite having a single-component disk.}
    \label{fig:edge_vc}
   
\end{figure*}

Our exercise indicates that the line-of-sight integration effect can significantly bias the deduced structure of the stellar disks. Detailed modeling provides a much clearer insight into the 3D distribution of stars in the stellar disks of galaxies.

As shown in Fig.~\ref{fig:fr}, the FR sharply peaks at $R \sim$ 4.4 kpc. This is due to the lower value of scale height, resulting from low velocity dispersion in the stellar disk. To check this further, we inspected the H$\alpha$ map of NGC~551 and found a maximum in the star formation activity around the same radius. It is possible that this young stellar population (as traced by H$\alpha$ emission) inherited the low stellar velocity dispersion from the cold gas, resulting in a lower scale height and higher FR values.

We also investigate the perceived signature of a multicomponent disk by examining the vertical cuts of the model edge-on surface brightness distribution of NGC~551. In Fig.~\ref{fig:edge_vc}, we plot vertical cut profiles at two different radii, i.e., 2.5 kpc and 3.5 kpc. We fit these vertical profiles with single (first and third panels) and double Gaussian (second and fourth panels) profiles. We compare the Bayesian information criterion (BIC) for the fits to determine which type of Gaussian provides a better fit. The BIC for the single Gaussian and double Gaussian are 786 and 534, respectively, indicating that a two-component fit is better. This result suggests the presence of a two-component disk in NGC~551, whereas the model disk has only one component (thick disk). Furthermore, as shown in the figure, the peak brightness of the thick disk in the double Gaussian fits (second and fourth panels) is higher than that of the thin disk component. This is not observed in galaxies. Therefore, decomposing the vertical cuts using multiple Gaussian components may not be a reliable method for investigating the presence of multicomponent stellar disks in galaxies.

\begin{figure}
        \includegraphics[width=\columnwidth]{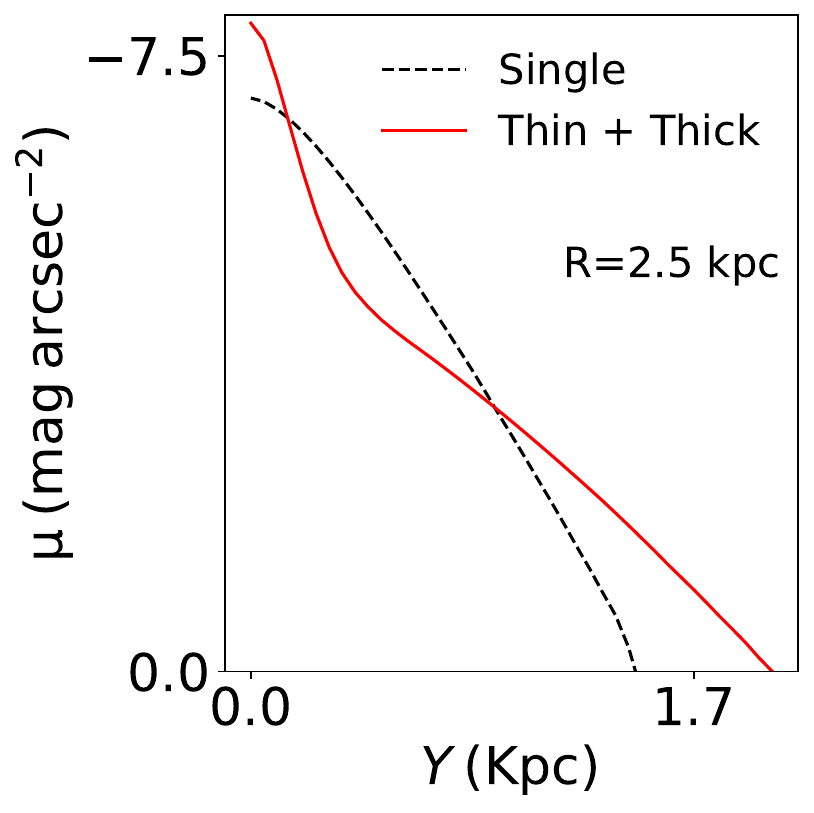}
    \caption{Vertical cuts of the simulated intensity distributions of the stellar disk in an edge-on orientation plotted on a logarithmic scale and at a radius of 2.5 kpc. The dashed black line represents the single-component disk, whereas the solid red line represents a two-component disk. An up-bending break appears only in the two-component disk.}
    \label{fig:log_scale_cut}
   
\end{figure}

In the literature, however, several studies have inspected vertical cuts of the edge-on surface density distribution in log scale to identify multicomponent disks \citep[see, e.g., ][]{2012ApJ...759...98C}. They detected an up-bending break in the log profile where a two-component disk model fits the observations better. To test this, we build a two-component model for the stellar disk in NGC~551. We chose a surface density ratio of the thick-to-thin disk ($f_\Sigma$) of 0.38, similar to that observed in the Milky Way \citep{2021MNRAS.507.5246M}. Following a dynamical modeling procedure similar to that used for the single-component disk, we build an edge-on surface density map for a two-component disk in NGC~551. In Fig. \ref{fig:log_scale_cut}, we plot the vertical cut profile of this map at a radius of 2.5 kpc on a logarithmic scale. For comparison, we also plot the same for the single-component disk. Both an up-bending and a down-bending break are shown for a two-component disk (solid red line), whereas no up-bending break is shown for a single-component disk (dashed black line). We note that the prominence of this up-bending depends on the surface density and velocity dispersion ratios of the thin and thick disks. For this study, we considered the disk of NGC~551 as a single-component one, which might have introduced some biases. However, we plan to investigate the suitability of a two-component stellar disk in NGC~551 in a future paper.

\section{Conclusions}

We assumed the galactic disk in NGC~551 to be a multicomponent system consisting of stellar, atomic, and molecular disks in vertical hydrostatic equilibrium under their mutual gravity in the external force field of the dark matter halo. We set up the corresponding joint Poisson's equations of vertical hydrostatic equilibrium and solved them numerically to obtain a 3D density distribution in the stellar disk. 

The stellar vertical velocity dispersion ($\sigma_\star$) plays a crucial role in determining the structure of the stellar disk under the vertical hydrostatic equilibrium. The $\sigma_\star$ decides the vertical pressure in the disk, which balances the gravity. Observationally determining $\sigma_\star$ is difficult. Many previous studies modeling galactic disks hydrostatically calculate $\sigma_\star$ analytically, considering the stellar disk to be an isolated single-component system in vertical hydrostatic equilibrium. However, in this study, for NGC~551, we used IFU observations from the CALIFA survey and estimated the $\sigma_\star$. 

The observed $\sigma_\star$ is the intensity-weighted velocity dispersion, MOM2, along a line-of-sight. We used an iterative method to consistently solve the hydrostatic equilibrium equations, beginning with MOM2 as the input to recover the correct $\sigma_\star$. At the end of this method, we recovered a $\sigma_\star$ profile, which consistently produces the observed MOM2 profile in a simulated observation. This velocity dispersion is found to be consistent with that calculated using the formalism of the DMS within $\sim 20\%$. However, this velocity dispersion significantly disagrees with the analytically calculated one described by Eq.~\ref{eq:analytical_expressiom}. For NGC~551, the intrinsic velocity dispersion can differ from the analytically calculated value by as much as 70\%.

The solutions of the hydrostatic equilibrium equation provide the 3D density distribution in the stellar, atomic, and molecular disks. For NGC~551, we solved the equations at $2 \leq R \leq 11$ kpc every 100 pc. We find that the vertical density distribution at all radii for all the components (stars, atomic, and molecular gas) considerably depends on the assumed $\sigma_\star$. Our estimated midplane density based on IFU data differs by $91\%$ for the stellar disk, and by $25\%$ and $22\%$ for the atomic and molecular disks, respectively, compared to the midplane densities calculated from the corresponding analytical expressions.

We estimated the scale heights (HWHM) in the stellar and gas disks in NGC~551 using the derived density solutions. We find the stellar scale height to vary between 200 - 470 pc. The stellar and gas disks are found to flare as a function of radius. We find that the analytically calculated $\sigma_\star$ overestimates the scale height in NGC~551 by about a factor of two. This is concerning, as it can artificially increase the thickness of the stellar disk by a large amount. Hence, using the correct stellar velocity dispersion, as estimated by spectroscopic observation, is critical in determining vertical scale height in stellar disks.

Using the density solutions and dynamical parameters of NGC~551, we built a 3D dynamical model of the stellar disk, which reproduces the observed MOM2 profile very well, indicating appropriate modeling of the intrinsic stellar velocity dispersion. At the same time, this modeling also reproduces the observed surface brightness distribution with minor deviations in the central region, likely due to non-axisymmetric structures such as spiral arms.

Next, we investigated how the stellar disk in NGC~551 would be observed in an edge-on orientation. We inclined our dynamical model to an inclination of $90\degree$ and projected it onto the sky plane. Using the traditional method adopted in the literature, we fit this edge-on intensity distribution with the product of the modified Bessel function of the first kind, $K_1(x)$ and $\rm sech^2$, as shown in Eq. \ref{eq:2d_fit_sech}. We find that this projection and fitting can artificially reduce the scale length of the stellar disk. Consequently, the estimated observed FR can also be decreased significantly. For NGC~551, the fitted FR is  $ 7.88 \pm 0.02$, whereas its actual ratio varies between 8 and 23 depending on the radius. Hence, estimating the structural parameters of stellar disks by simply 2D-fitting the edge-on surface density might be affected by line-of-sight integration effects.

\begin{acknowledgements}

We are grateful to the anonymous referee for the useful comments and suggestions, which substantially helped improve the content of our article. HR thanks Karl M. Menten for supporting this study through granting an MPIfR guest-research fellowship. We thank Nadezda Tyulneva for her contribution in the development of the MCMC NFW dark matter halo density profile model code used in this study.

We thank the staff of the GMRT that made these observations possible for obtaining the \HI~data of galaxy NGC~551 (code $39\_037$) as part of the “Mass modelling and Star Formation Quenching of Nearby Galaxies” (MasQue; PI: V. Kalinova) project.

This study uses data provided by the Calar Alto Legacy Integral Field Area (CALIFA) survey\footnote{\url{https://califa.caha.es/}}, based on observations collected at the Centro Astronómico Hispano Alemán (CAHA) at Calar Alto, operated jointly by the Max-Planck-Institut für Astronomie and the Instituto de Astrofísica de Andalucía (CSIC). 
This work extensively uses data from the EDGE-CALIFA \citep{2017ApJ...846..159B}. 
HR appreciates the CALIFA and EDGE-CALIFA teams for making the data accessible to the general public. 
NP acknowledges support from the Department of Science and Technology (DST), Government of India, through Startup Research Grant (Project no: SRG/2022/00917). 
DC acknowledges support by the Deut\-sche For\-schungs\-ge\-mein\-schaft, DFG\/ project number SFB1601 B3. 
SFS thanks the PAPIIT-DGAPA AG100622 project and CONACYT grant CF19-39578. 
This work was supported by UNAM PASPA–DGAPA. 
V.\,V. acknowledges support from the ALMA-ANID Postdoctoral Fellowship under the award ASTRO21-0062.

\end{acknowledgements}

\bibliographystyle{aa}
\bibliography{ref}

\begin{appendix}

\section{Kinematic modeling}
\label{sec:kin_mod}
\begin{figure}
    \centering
     \begin{subfigure}[b]{0.42\textwidth}
         \centering
         \includegraphics[width=\textwidth]{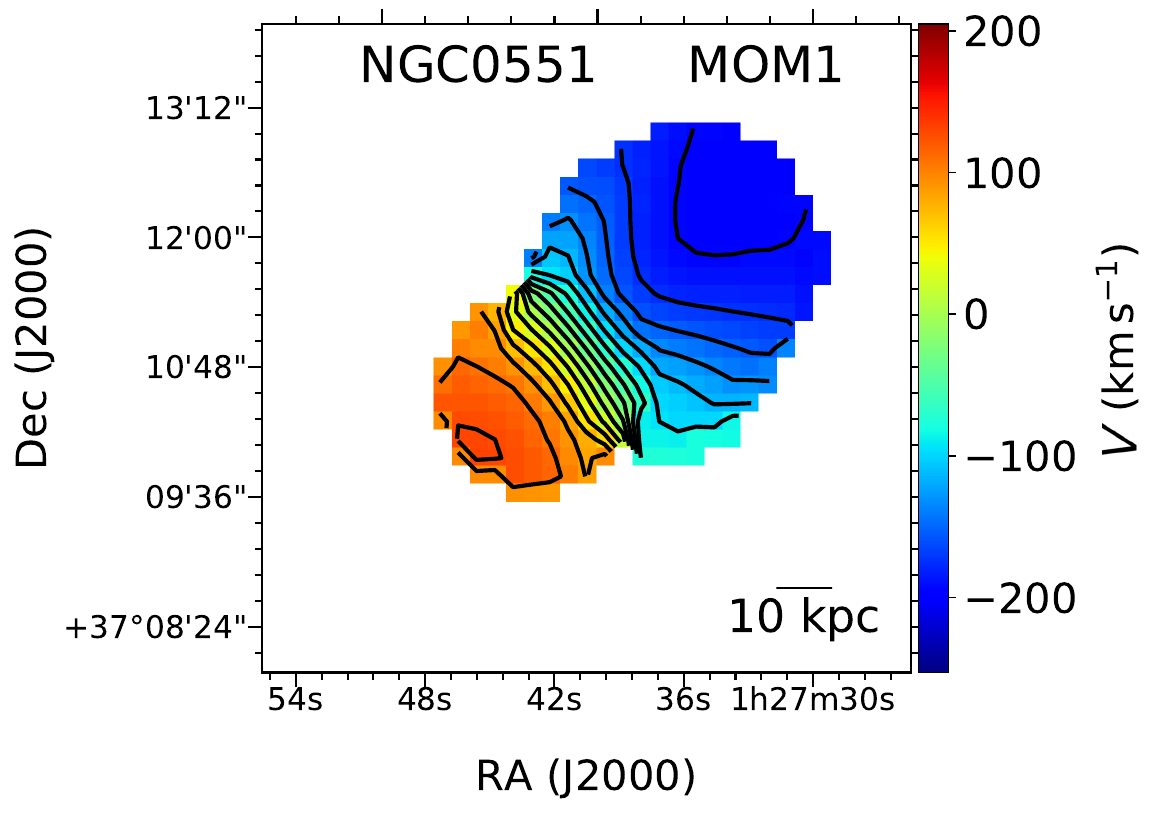}
         \caption{}
         \label{fig:app_mom1}
     \end{subfigure}
     \begin{subfigure}[b]{0.4\textwidth}
         \centering
         \includegraphics[width=\textwidth]{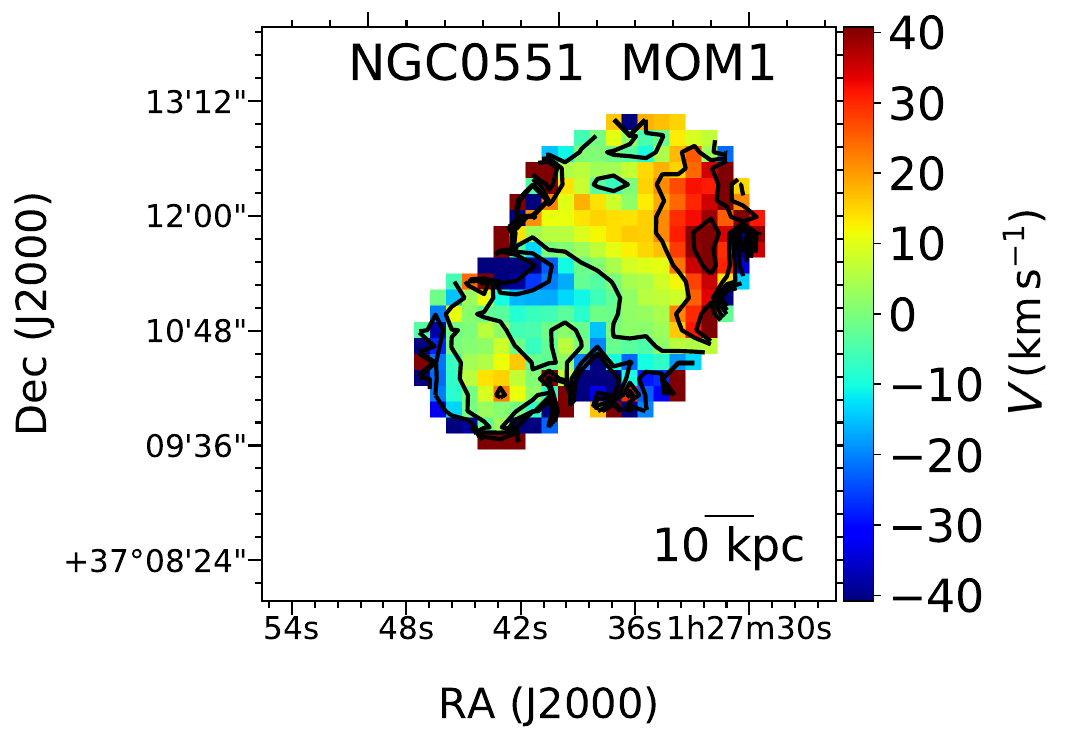}
         \caption{}
         \label{fig:mom1_red}
     \end{subfigure}
     \caption{ Panel a: \HI~Moment one map. The black contours are drawn at 20 \kms~intervals. Panel b: Moment one residual map, i.e., the difference between the moment one map obtained from data and that from the model.}
     \label{fig:mom_comp}
\end{figure}

As mentioned in Sect.~\ref{subsec:rotc}, the kinematic modeling and, hence, the rotation curve of NGC~551 have been derived by fitting the 3D tilted ring model to the \HI~spectral cube using FAT \citep{2015MNRAS.452.3139K} following the procedure mentioned in \citet{2023MNRAS.524.6213B}. To check the quality of the fit, we look for the moment one residual map that represents the difference between the observed velocity field and the modeled velocity field. Figure \ref{fig:mom_comp} shows the moment one map derived from observed data and the corresponding residual map obtained after the kinematic modeling. Figure \ref{fig:mom1_red} shows that $\sim 80\%$ of the residual velocity lies within  $\pm30$ \kms, while the rotation velocity goes much higher than that (in the central region $160$ \kms~and in the outer region $180$ \kms). This ensures the kinematic modeling was done adequately. In future work, for some galaxies, we will be able to improve our mass modeling technique further, combining the circular velocity curves (CVC) of CALIFA galaxies from different galaxy components (Kalinova, Tyulneva, et al., in prep; Biswas et al., in prep.). For a subsample of CALIFA galaxies (for those systems where the different datasets are available), we will be able to combine the inner CVC of the galaxies from stellar dynamics \citep{2017MNRAS.469.2539K} and the inner CVC from molecular gas kinematics \citep{2018ApJ...860...92L} with their outer CVC from atomic gas \citep{2023MNRAS.524.6213B} to model their gravitational potential. 

\section{Mass modeling}
\label{sec:mass_mod_app}
\begin{figure}
        \includegraphics[width=0.45\textwidth]{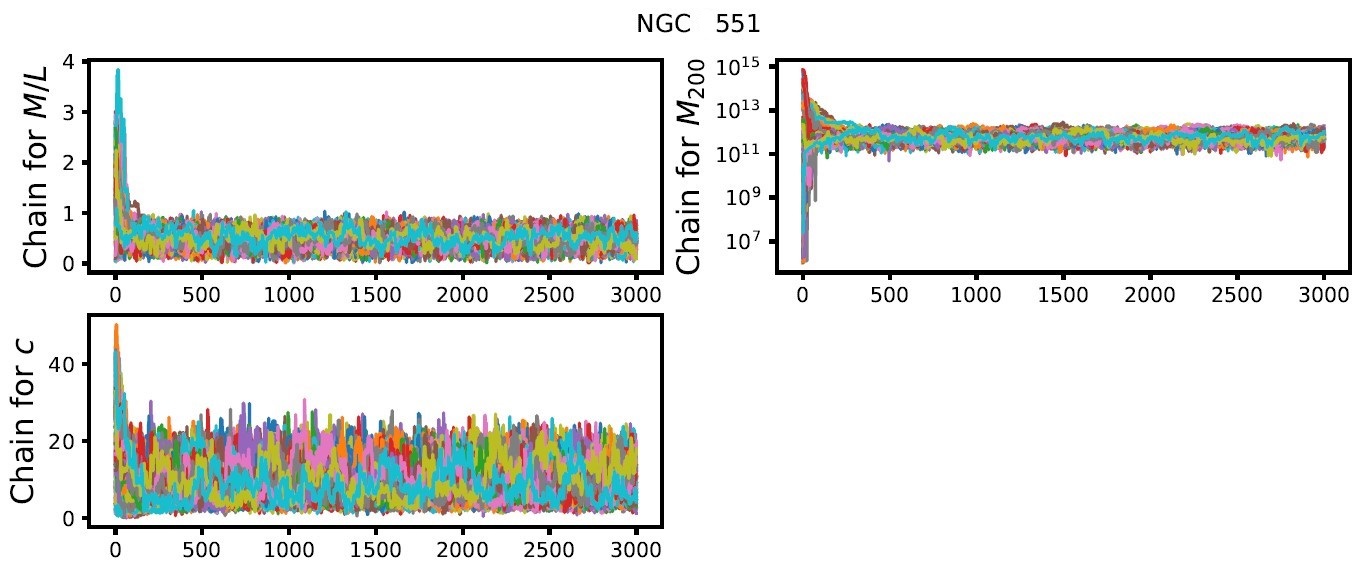}
    \caption{Distribution of the Burn-in chains during the MCMC procedure.}
    \label{fig:burning_chainf}
\end{figure}

\begin{figure}
        \includegraphics[width=0.45\textwidth]{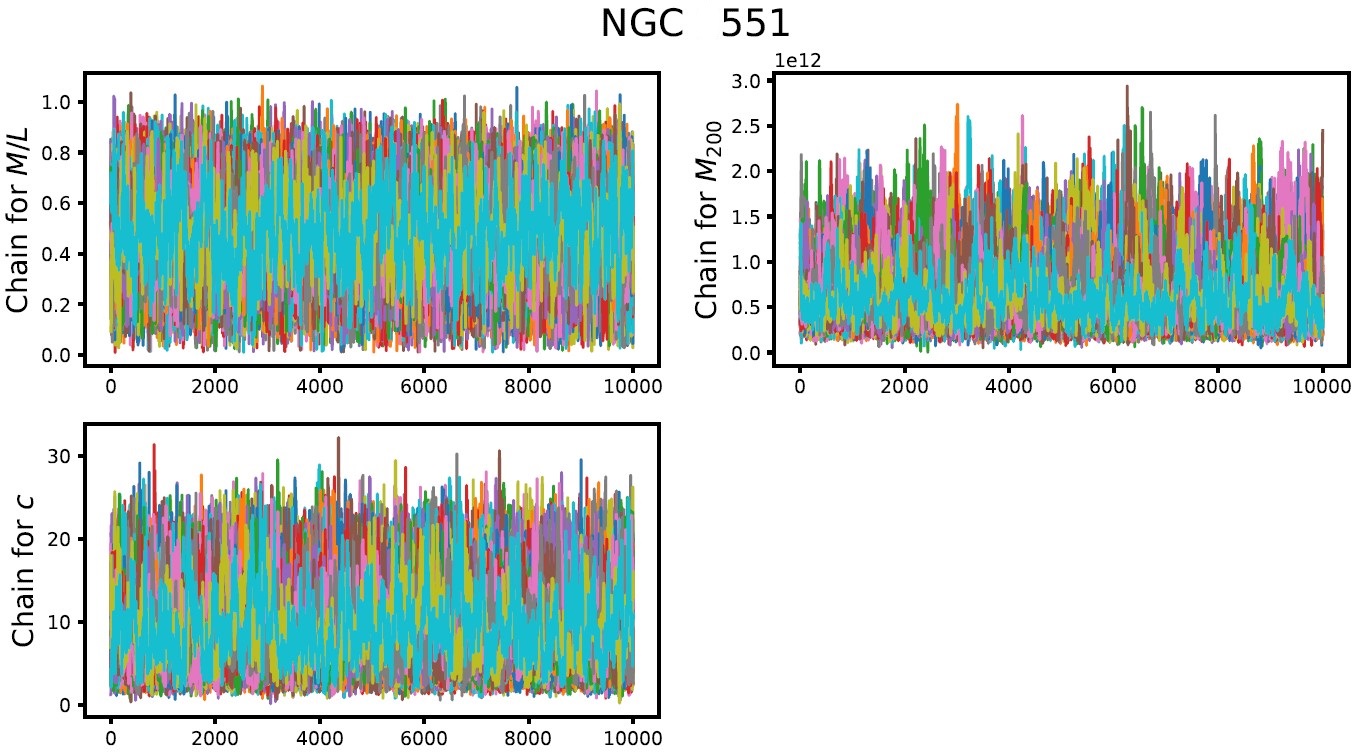}
    \caption{Distribution of the posterior chains during the MCMC procedure.}
    \label{fig:final_chainf}
\end{figure}

\begin{figure}
        \includegraphics[width=0.45\textwidth]{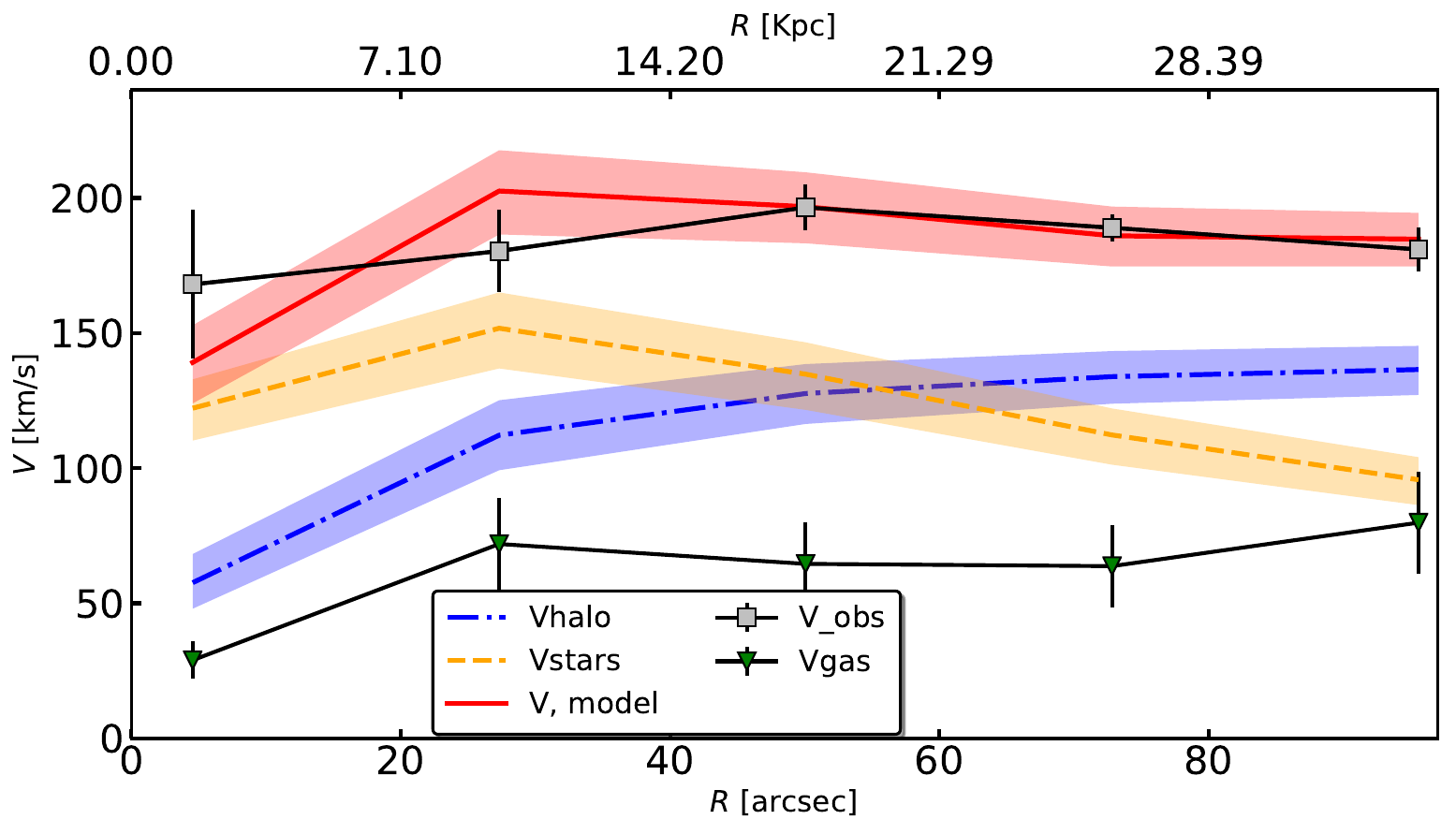}
    \caption{Modeled rotation curve along with the contributions of star, gas, and halo for the source NGC~551. }
    \label{fig:mcmcf}
\end{figure}

\begin{figure}
        \includegraphics[width=0.45\textwidth]{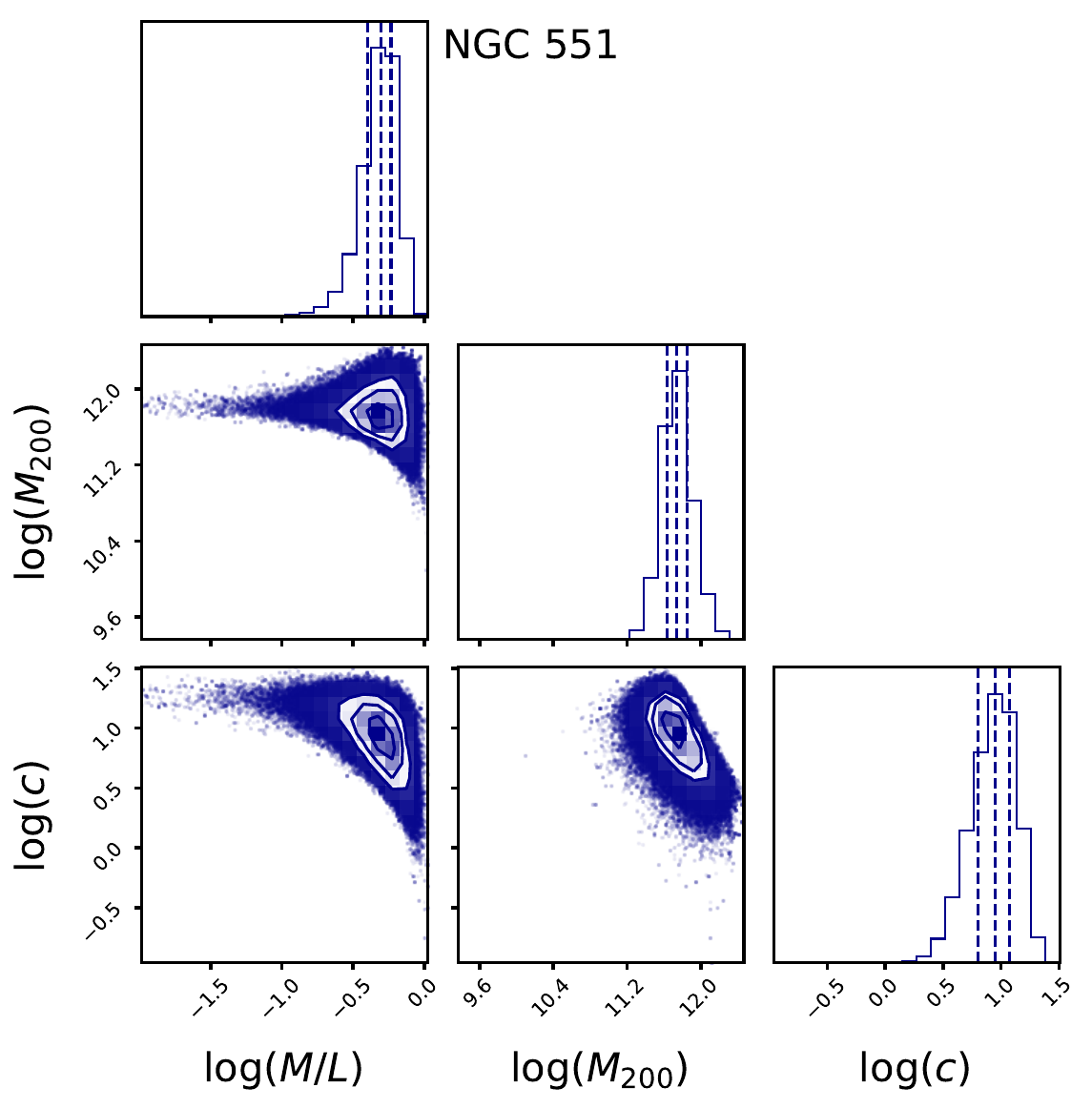}
    \caption{Posterior distribution of the free parameters (i.e., $M/L$, $M_{\rm 200}$ and $C$) used for modeling. }
    \label{fig:cornerf}
\end{figure}

As discussed in Sect.~\ref{sec:dm}, to obtain $V_{\text{halo}}$, mass modeling of the galaxy has been performed using the MCMC optimization method. As our primary interest is finding the halo parameters, we combined the \HI~and H$_2$ surface densities to get a gas surface density for simplicity. We multiplied the \HI~surface density by a factor of 1.4 to account for helium. Assuming the gas disk to be thin, we calculated its circular velocity analytically \citep[see Eq.~5 of ][]{2023MNRAS.524.6213B}. The total rotation velocity can be written as follows:
\begin{equation}
     V_{\rm rot}^2 = V_{\rm star}^2 + V_{\rm gas}^2 + V_{\rm halo}^2,
\end{equation}
\noindent where $V_{\rm star}$, $V_{\rm gas}$ and $V_{\rm halo}$ are the stellar, total gas, and the dark matter halo component of the rotation velocity, respectively. We used a three-parameter model to fit the observed rotation curve. One of the free parameters in our model is the stellar $M/L$, i.e., the mass-to-light ratio, which is kept constant with radius. This same stellar $M/L$ was used while estimating the stellar velocity component from stellar surface brightness using JAM \citep{2008MNRAS.390...71C}. As mentioned earlier, we considered an NFW profile representing the dark matter halo. For a halo density  given by Eq.~\ref{eq:equation 8}, the velocity component of the halo can be expressed through the following equation \citep[see][for details]{2023MNRAS.524.6213B}: 

\begin{align}
\begin{split}
         V_{\rm halo} (R) = 0.014 M_{\rm 200}^{1/3} &\sqrt{\frac{20.24 M_{\rm 200}^{1/3}}{R}}\times \\  
        & \sqrt{\frac {\ln \Bigl(1+ \frac{RC}{20.24 M_{\rm 200}^{1/3}}\Bigr) \frac{RC/(20.24 M_{\rm 200}^{1/3})}{1+ RC/(20.24 M_{\rm 200}^{1/3})} } {\ln({1+C}) - \frac{C}{1+C}} },
    \end{split}
\end{align}

\noindent where $M_{\rm 200}$ is the halo mass of the galaxy. This is measured as the dark matter mass enclosed inside a sphere having the same center as the galaxy and a radius where the dark matter density falls to 200 times the critical density of the Universe. $ C = R_{\rm 200}/R_{\rm s}$ represents the concentration parameter of the halo. Thus,  $ M_{\rm 200}$ and $C$ are the other two free parameters of our model. We used a similar procedure as mentioned in \citet{2023MNRAS.524.6213B} for implementing the MCMC optimization, i.e., the selection of the prior is motivated solely from the previous independent observations and is not dependent on any of the previous scaling relations (e.g., stellar mass versus halo mass relation \citep{2013MNRAS.428.3121M}, $M_{\rm 200}$ vs. $C$ relation \citep{2008MNRAS.391.1940M,2014MNRAS.441.3359D} etc.). The details of the prior selection are mentioned elaborately in Sect.~ 5 of \citet{2023MNRAS.524.6213B}. We used the same set of priors as they work adequately for galaxies with different morphologies and kinematics. For NGC~551, we used 130 walkers, 3000 burn-in steps, and 10000 posterior steps to fit the model through the MCMC optimization procedure. Figures \ref{fig:burning_chainf} and \ref{fig:final_chainf} show the distribution of the burn-in and posterior chains during the MCMC run, respectively. From Fig. \ref{fig:burning_chainf}, we see that the burn-in chains indeed reach the stationary phase with a well-sampled posterior parameter space. Figure~\ref{fig:mcmcf} shows the mass modeling of NGC~551 through the MCMC optimization, whereas Fig.~\ref{fig:cornerf} presents the posterior distribution of the free parameters. As shown in these plots, the mass modeling for NGC~551 has been done satisfactorily. Furthermore, we found the MCMC optimized value for $M/L$ is $0.53^{+0.14}_{-0.15}$, which is consistent with that obtained using the calibration of \citet{2012AJ....143..139E}, i.e., 0.51. This demonstrates the consistency of our mass modeling, and hence the halo parameters found in this process can be used reliably for hydrostatic modeling of the stellar disk. 

\section{Derivation of the disk scale length through MGEFIT}
\label{sec:scale_length}

\begin{figure}
        \includegraphics[width=\columnwidth]{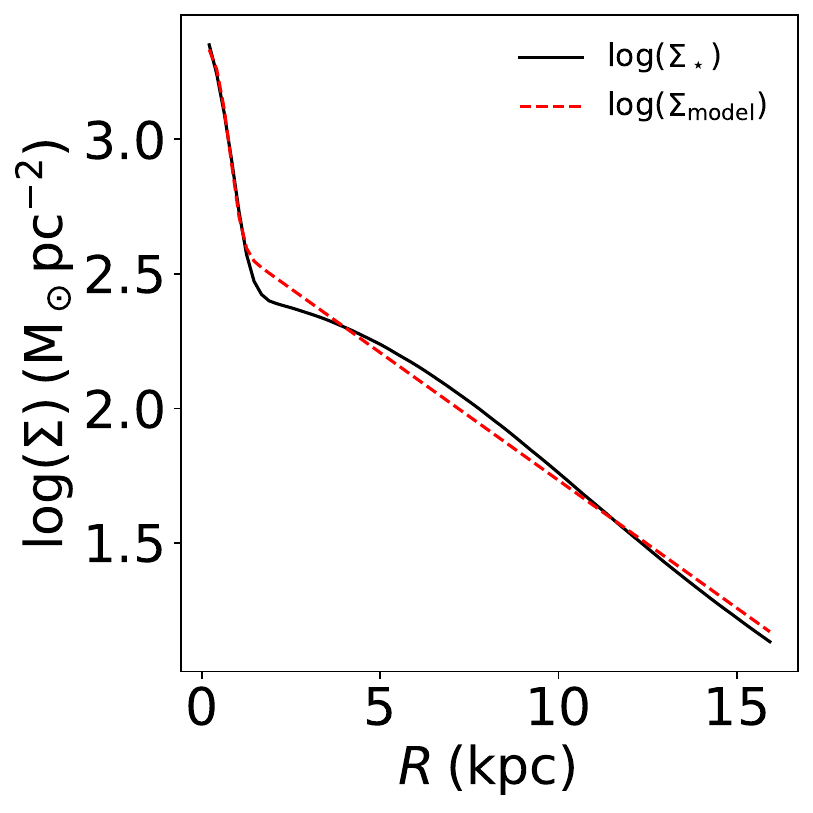}
    \caption{1D fit to the log of stellar surface density profile. The solid black line represents the stellar surface density profile (as obtained by MGE fit), and the dashed red line represents the combined fit of a Sérsic bulge and an exponential disk component.}
    \label{fig:surf_fit}
\end{figure}

We described the MGE fit procedure in Sect.~\ref{sec:surf}. Here, we present the model fitting to the surface density profile. We modeled the galaxy light distribution with a combination of a Sérsic bulge and an exponential disk component described by:
\begin{equation}
\Sigma(r) = \Sigma_{\rm e} \exp\left[
- \kappa\left[ \left( \frac{r}{r_{\rm e}} \right)^{\frac{1}{n}} - 1 \right]
\right] + \Sigma_{\rm o} \exp\left( -\frac{r}{r_{\rm s}} \right) ,
\label{eq:ser_exp}
\end{equation}
where $\Sigma_{\rm e}$ is the surface brightness of the Sérsic component at the effective (half-light) radius $r_{\rm e}$. $n$ is the Sérsic index, which controls the curvature/shape of the Sérsic component. $\kappa$ is a dimensionless constant that depends on $n$. $\Sigma_{\rm o}$ is the central surface brightness of the exponential disk component, whereas $r_{\rm s}$ is the exponential scale length of the disk component. For the value of $\kappa$, we used the equation
\begin{equation*}
\kappa = 2\,n - \frac{1}{3} + \frac{4}{405\,n} + \frac{46}{25515\,n^2},
\end{equation*}
where $n$ is the Sérsic index \citep{1999A&A...352..447C}. We performed the fit in the logarithmic domain. In this approach, both the observed stellar surface density data and the model given by Eq.~\ref{eq:ser_exp} were transformed using the logarithm (i.e., we fit $\log(\Sigma_\star)$ versus $\log(\Sigma_{\rm model})$). This procedure is advantageous when dealing with data spanning several orders of magnitude, as it emphasizes relative differences across the dynamic range. We performed the fit using the \texttt{lmfit} package in Python. The best-fit parameters obtained from the fit, along with their associated uncertainties, are summarized in Table~\ref{tab:mgefit_params}. In Fig.~\ref{fig:surf_fit}, we show the fit to the observed surface density profile.  

\begin{table}
\caption{Best-fit parameters for a Sérsic and exponential model}
\label{tab:mgefit_params}
\centering
\begin{tabular}{lcc}
\hline\hline
Parameter & Value \\
\hline
$\rm \Sigma_{\rm e}\,\:(M_\odot\:pc^{-2})$  & $ 1090 \pm 120$ \\
$\rm \Sigma_{\rm o}\,\:(M_\odot\:pc^{-2})$  & $481 \pm 13$ \\
$r_{\rm e}\,\:\rm (pc)$  & $543 \pm 24$ \\
$r_{\rm s}\,\: \rm (pc)$ & $4570 \pm 60$     \\
$n$ & $0.39 \pm 0.08$    \\
\hline
\end{tabular}
\end{table}

\end{appendix}
\end{document}